\documentclass[a4paper,12pt]{article}
\usepackage[utf8]{inputenc}
\usepackage[english]{babel}
\usepackage[width=0cm, left=1in, right=1in, top=1in, bottom=1in]{geometry}
\usepackage{setspace}
\usepackage{amsmath}
\usepackage{amsfonts}
\usepackage{amssymb}
\usepackage{graphicx}
\usepackage{enumerate}
\usepackage{enumitem}
\usepackage{array}
\usepackage{lscape}
\usepackage{authblk}
\usepackage{multicol}
\usepackage{adjustbox}
\usepackage{csquotes}
\usepackage{float}
\usepackage{graphicx,subcaption} 
\usepackage{sectsty}
\sectionfont{\fontsize{14}{15}\selectfont}
\subsectionfont{\fontsize{14}{15}\selectfont}
\usepackage{ragged2e}
\usepackage{mathtools}  
\newtagform{dots}{\ldots(}{)}
\numberwithin{equation}{section}
\counterwithin{figure}{section}
\counterwithin{table}{section}

\usepackage{booktabs}
\usepackage{authblk}\usepackage{openbib}\usepackage{epsf}\marginparwidth 0pt \oddsidemargin 0pt \evensidemargin 0pt \marginparsep 0pt \topmargin 0pt \topmargin .5cm \headheight 0pt \headsep 0pt
\usepackage{booktabs}\usepackage{dcolumn}\usepackage{caption}
\usepackage{threeparttable}\usepackage{booktabs}
\usepackage{rotating}\usepackage{booktabs}\usepackage{ulem}
\usepackage{indentfirst}\usepackage{array}\usepackage{times}\usepackage{color}\usepackage{psfrag}\usepackage{eso-pic}
\usepackage{type1cm}
\date{}
\title{ \bfseries\Large A REVISIT TO MAXIMUM LIKELIHOOD ESTIMATION OF WEIBULL MODEL PARAMETERS}
\author[1]{Buu-Chau Truong}
\author[2]{Peter Mphekgwana}
\author[1,3*]{Nabendu Pal}
\affil[1]{\normalsize Faculty of Mathematics and Statistics, Ton Duc Thang University, Ho Chi Minh City, Vietnam}
\affil[2]{\normalsize  Department of Research Administration and Development, University of Limpopo, \qquad  South Africa}
\affil[3]{\normalsize  Department of Mathematics, University of Louisiana at Lafayette, USA}
\affil[*]{\normalsize  Corresponding author: \it{nabendu.pal@tdtu.edu.vn}; \it{\normalsize nabendu.pal@louisiana.edu}; \it{\normalsize  nabendupal@gmail.com}}
\date{}

\begin{document}
\maketitle \setcounter{page}{1}
\begin{abstract}
In this work, we revisit the estimation of the model parameters of a Weibull distribution based on \textit{iid} observations, using the maximum likelihood estimation (MLE) method which does not yield closed expressions of the estimators. Among other results, it has been shown analytically that the MLEs obtained by solving the highly non-linear equations do exist (i.e., finite), and are unique. We then proceed to study the sampling distributions of the MLEs through both theoretical as well as computational means. It has been shown that the sampling distributions of the two model parameters' MLEs can be approximated fairly well 
by suitable Weibull distributions too. Results of our comprehensive simulation study corroborate some recent results on the first-order bias and first-order mean squared error (MSE) expressions of the MLEs.   \\  

\end{abstract}

\section{Introduction}
The main purpose of this work is to provide some interesting insights of the sampling distributions of the MLEs of the model parameters of a two-parameter Weibull distribution. Even though a good amount of work has been done to study the bias and the MSE of the aforementioned MLEs, both computationally as well as theoretically, their non-asymptotic sampling distributions have remained elusive in the literature, and this work is going to shed some new light on this topic. This section has been divided as follows: In  Subsection 1.1 we first review the basic properties of Weibull distribution very briefly including a list of wide variety of applications. Subsection 1.2 gives a summary of recent works on bias and MSE of the MLEs. Note that our objective is much brother; - to tackle the sampling distributions of the MLEs for small to moderate sample sizes, and then the bias and the MSEs come as products which are found to be matching with the ones reported already in the literature fairly well.
\subsection {Preliminaries} 
\begin{figure}[H]
\includegraphics[width=15cm,height=9.5cm]{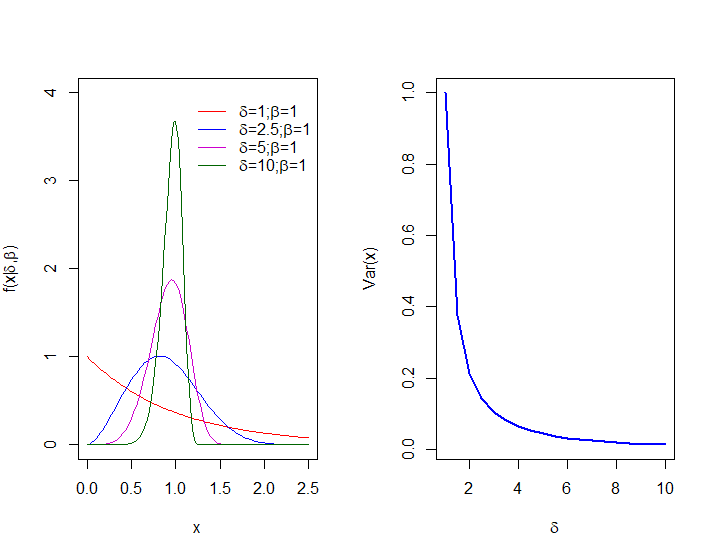}
\centering
\caption{Plots of $W(\delta, \beta)$ \textit{pdfs} and its variance for various $\delta$, with fixed $\beta =1$}
\label{fig1}
\end{figure}

A two-parameter Weibull distribution, henceforth denoted by $W(\delta,\beta)$, is characterized by its probability destiny function (\textit{pdf}) given as  
\setlength{\belowdisplayskip}{1.5pt} \setlength{\belowdisplayshortskip}{1.5pt}
\setlength{\abovedisplayskip}{1.5pt} \setlength{\abovedisplayshortskip}{1.5pt} 
\begin{equation}
f(x|\delta,\beta) = \left( \delta/\beta \right) \left( x/ \beta \right)^{\delta-1} \textit{exp} \{ - \left( x/ \beta \right)^\delta \}, x>0;
\end{equation}
where $\delta >0$ and $\beta >0$ are the shape as well as scale parameters, respectively. The corresponding cumulative distribution function (\textit{cdf}) of $W(\delta, \beta)$ is
\setlength{\belowdisplayskip}{1.5pt} \setlength{\belowdisplayshortskip}{1.5pt}
\setlength{\abovedisplayskip}{1.5pt} \setlength{\abovedisplayshortskip}{1.5pt} 
\begin{equation}
F(x|\delta,\beta) = 1-\textit{exp}  \{ - \left( x/ \beta \right)^\delta \}, x>0;
\end{equation}
and $0$ for $x \le 0$. For any constant $r$, define $E(X^r) = m_r$. When $X$ follows $W(\delta,\beta)$, it is easy to show that
\setlength{\belowdisplayskip}{1.5pt} \setlength{\belowdisplayshortskip}{1.5pt}
\setlength{\abovedisplayskip}{1.5pt} \setlength{\abovedisplayshortskip}{1.5pt} 
\begin{equation} \label{eq:moment}
m_r = \beta^r \Gamma \left(1+ r/ \delta \right),
\end{equation}
provided \eqref{eq:moment} exists. It is trivial to see that $\mu = E(X)= m_1 =\beta \Gamma \left( 1 + 1/\delta \right)$ and $\sigma^2 = \text{Var}(X) = \beta^2 \left( \Gamma \left(1+2/\delta \right) - \left\lbrace \Gamma \left( 1 + 1/\delta \right) \right\rbrace^2 \right)$, where for any constant $k$,
\begin{equation}
\Gamma (k) = \int_{0}^{\infty} x^{k-1} \textit{exp} (-x)  d{x}  
\end{equation}
 is the usual gamma function. Unlike the gamma model which increases stochastically with respect to its shape parameter, Weibull behaves differently. For fixed $\beta$, as $\delta$ increases, $W(\delta,\beta)$ becomes more symmetric and concentrated near $\beta$. This can be seen from Figure \ref{fig1}. Again, while a gamma model always has a non-negative measure of skewness w.r.t its shape parameter, Weibull's behavior is quite different. The measure of skewness of $X \sim W(\delta,\beta)$ is 
\setlength{\belowdisplayskip}{1.5pt} \setlength{\belowdisplayshortskip}{1.5pt}
\setlength{\abovedisplayskip}{1.5pt} \setlength{\abovedisplayshortskip}{1.5pt} 
\begin{equation}
\gamma_1 (\delta) = E((X-\mu)/\sigma)^3 = \left\lbrace 2\Gamma_1^3-3\Gamma_1\Gamma_2+\Gamma_3\right\rbrace / \left\lbrace \Gamma_2- \Gamma_1 \right\rbrace ^ {3/2},
\end{equation}
where $\Gamma_i = \Gamma(1+i/\delta)$, $i=1,2,3$. The following Figure \ref{fig2} shows the plot of $\gamma_1 (\delta)$ against $\delta$, and as $\delta$ exceeds $\delta_0 = 3.6$, the distribution turns negatively skewed from being positively skewed.

Applications of the $W(\delta,\beta)$ model in real-life problems cannot be overemphasized. From engineering reliability and life testing to nuclear physics and astronomy, this model has been widely used. For a general overview of the distribution including its properties see Johnson, Kotz and Balakrishnan (1994). For other general usage of this model in reliability studies one can see Xie, Lai and Murthy (2003); Murthy, Bulmer and 
Eccleston (2004) and the other relevant references therein.

 Lately, Weibull distribution is getting renewed interest due to its applications in renewable energy research, 
especially in wind energy generation. In this regard two interesting studies are being cited: Aljeddani and Mohammed (2023) as well as Aziz et al. (2023). For some other specific and interesting applications, one can see (i) Chandrasekhar (1943), Eliazar (2017) (in Physics and Astronomy); (ii) Sharif and Islam (1980), Chatfield and Goodhardt (1973) (in Tech and Business Management); (iii) Austin et al. (1984), Almeida (1999), Fok et al. (2001), Tadikamalla (1980), Keshavan et al. (1980), Sheikh et al. (1990) (in Industrial Engineering); (iv) Collett (2015) (in Medical Research); (v) Heo, Boes and Salas (2001) (in Hydrology); (vi) Mafart et al. (2002) (in Food Science); (vii) Fleming (2001) (in Ecology); (viii) Carroll (2003) (in Clinical Trials); (ix) Li et al. (2003) (in Civil Engineering); and (x) Matsushita et al. (1992) (in Epidemiology). Most of these aforementioned applications are based on uncensored data. Also, between the two Weibull parameters, it is the shape parameter ($\delta$) that carries more importance and 
has profound implications when $W(\delta,\beta)$ is used to model a real-life dataset, especially in the context of engineering problems. For a comprehensive review on this topic see the excellent paper by Jiang and Murthy (2011).

 \begin{figure}[H]
\includegraphics[width=15cm,height=9.5cm]{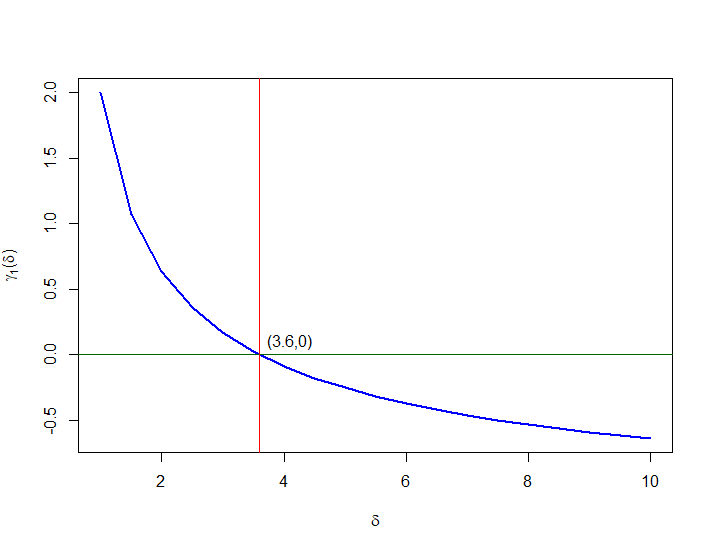}
\centering
\caption{Weibull measure of skewness $\gamma_1 (\delta)$ as a function of $\delta$}
\label{fig2}
\end{figure} 

Even though many generalizations of $W(\delta,\beta)$  have been proposed in recent years, Weibull still remains as one of the central figures in statistical modelling and methods due to its natural appeal; and the simplistic nature of its \textit{cdf} allows one to study its quantiles with relative ease. This is the main motivation behind this revisit to $W(\delta,\beta)$ and filling certain gaps analytically as well as computationally while estimating the model parameters. 

\subsection{Some Recent Results.}
Based on i.i.d observations $X_1, X_2,\dots, X_n$ (= \textbf{X}, say) form $W(\delta,\beta)$, the log-likelihood function is
\setlength{\belowdisplayskip}{1.5pt} \setlength{\belowdisplayshortskip}{1.5pt}
\setlength{\abovedisplayskip}{1.5pt} \setlength{\abovedisplayshortskip}{1.5pt} 
\begin{equation}
L_*(\delta,\beta|\textbf{X}) = n \textit{ln} \delta - (n\delta) \textit{ln} \beta + n (\delta -1) \textit{ln} \tilde{X} - \textstyle \sum_{i=1}^{n} (X_i/\beta)^\delta,
\end{equation}
where $\tilde{X}= (\prod_{i=1}^{n} X_i )^{1/n}$ is the geometric mean (GM) of the observations. The MLEs of $\delta$ and $\beta$ are found in the following way. First solve $h(\delta|\textbf{X})=0$ to obtain $\hat{\delta}(=\hat{\delta}_{MLE})$, where
\setlength{\belowdisplayskip}{1.5pt} \setlength{\belowdisplayshortskip}{1.5pt}
\setlength{\abovedisplayskip}{1.5pt} \setlength{\abovedisplayshortskip}{1.5pt}
\begin{equation}\label{h_function}
h(\delta |\textbf{X}) = (1/\delta)+ \textit{ln} \tilde{X} - \{ \textstyle \sum_{1}^{n} X_{i}^{\delta} (\textit{ln} X_i) \} / \{\textstyle \sum_i^n X_i^\delta \}.
\end{equation}
Then, the estimator of $\beta$, say $\hat{\beta}  (=\hat{\beta}_{MLE})$, is obtained as 
\begin{equation}
\hat{\beta} = \{ \textstyle \sum_1^n X_i^{\hat{\delta}} / n \} ^{1/\hat{\delta}}.
\end{equation}
Even though the above MLEs do not have explicit closed expressions, a lot of studies have been carried out to study the bias and the mean squared error (MSE) of $\hat{\delta}$ and $\hat{\beta}$ mainly through numerical computations. It is known that $\hat{\delta}$ tends to over estimate $\delta$, i.e., $\hat{\delta}$ is positively biased, whereas $\hat{\beta}$ is not so. For \enquote {small} values of $\delta$, $\hat{\beta}$ is positively biased and the opposite happens when $\delta$ is not \enquote{small}. For more on this matter, one can see Tanaka et al. (2018) some of which are given as discussed below.

The first order bias expression of $\hat{\delta}$ is a function of $\delta$ only, apart from being dependent on $n$. On the other hand, the first order bias of $\hat{\beta}$ is a function of $\delta$ multiplied by $\beta$ which can be seen from the following results. 

In a major work on Weibull parameters' estimation, from a `second order' decision-theoretical point of view, Tanaka et al. (2018) derived, among other useful results, the following asymptotic first order bias expressions for $\hat{\delta}$ and $\hat{\beta}$ as 
\begin{equation}\label{1_orderbias_delta}
\begin{split}
E(\hat{\delta}-\delta) = (B_1/n)  + o(1/n),\\
E(\hat{\beta}-\beta) = (B_2/n)  + o(1/n);
\end{split}
\end{equation}
where 
\begin{equation} \label{B1}
	B_1 = \delta (3 \zeta(2) - \zeta(3))/\zeta^2(2)\simeq \delta (1.37953),
\end{equation}
and 
\begin{equation} 
 	B_2=\beta \{(A_1^2+\zeta(2))\zeta(2)-\delta\{ 2\zeta^2(2)-(4A_1+1)\zeta(2)+2A_1\zeta(3)\} / (2\delta^2\zeta^2(2) \},
\end{equation}
	   with $A_1=1-\gamma$, $\gamma=0.577216\dots$ is the Euler's constant, and $\zeta(z)= \textstyle \sum_{i=1}^\infty (1/i^z), z>1$, is the Riemann zeta function. Further simplification shows that, 
\begin{equation}\label{E_beta}
E((\hat{\beta}/ \beta) - 1) \simeq \ [ (1/\delta^2)(0.554332\dots)-(1/\delta)(0.369815\dots)\ ] /n.
\end{equation}
Thus, the above expressions show that $\hat{\delta}$ has a positive asymptotic bias. However, $\hat{\beta}$ is not always asymptotically positively biased, and it depends on the value of $\delta$. For ``small" values of $\delta$ (i.e., $0<\delta<1.4989 \approx 1.5)$, $\hat{\beta}$ is positively biased, and for ``large" values of $\delta$ (i.e., $\delta >1.4989 \approx 1.5$), $\hat{\beta}$ is negatively biased.

Two recent works are worth mentioning. Chen et al. (2017) obtained bias corrected modified versions of the MLEs, and compared them in terms of bias and MSE with the actual MLEs for small to moderately large sample sizes through simulation. This bias correction was done by first order cumulant expansions of the MLEs put forward by Cox and Snell (1968). However, it should be noted that exactly same was done by Tanaka et al. (2018) to study such estimators' second order decision-theoretic properties analytically based on an extension of the result by Karlin (1958) provided by Takeuchi and Akahira(1979). In a more recent study Makalic and Schmidt (2023) extended the bias correction of the Weibull MLEs for censored samples. But notice again that these works are about studying the bias and MSE of the MLEs, not for their sampling distributions which is the focus of our investigation.

In this work, we are focusing solely on the maximum likelihood estimators of the model parameters which
are to be obtained by solving highly nonlinear equations. Though there are other easy to implement estimation techniques 
available (such as the method of moments estimation), those techniques tend to cause some loss of information due to not 
using the (minimal) sufficient statistics. For a complete review of various estimation techniques (other than the
MLE) with complete as well as censored data one can refer to the book by McCool (2012).

\section{Two Analytical Results} 
The fact that the first order bias expression of $\hat{\delta}$ and $(\hat{\beta}/\beta)$ see (\eqref{1_orderbias_delta} – \eqref{E_beta}) are functions of $\delta$ only (apart from $n$) shouldn't come as a surprise as the following result shows.\\

\noindent {\bf{Result 2.1}:} The probability distributions of $\hat{\delta}$ and $(\hat{\beta}/\beta)$ (i.e., the sampling distributions of the MLEs) are dependent only on $\delta$, apart from the sample size $n$.

The proof of the above results is easy once the transformation $Y_i = (X_i/\beta), 1 \leq i \leq n$, is used. Note that $Y_i$'s are i.i.d $W(\delta,1)$, and $h(\delta |\textbf{X})= h(\delta |\textbf{Y})$, and hence the solution $\hat{\delta}$ of $h(\delta |\textbf{X})=0=h(\delta |\textbf{Y})$ has a distribution that is dependent on $\delta$ only. At the same time, the scaled version of $\hat{\beta}$ depends only on $Y_i's$, i.e., $(\hat{\beta}/ \beta)= \{ \textstyle \sum_1^n Y_i^{\hat{\delta}} / n \} ^{1/\hat{\delta}}$, thereby making the distribution of $(\hat{\beta}/ \beta)$ dependent on $\delta$ only.
 
However, the existing literature does not say clearly whether a solution to the equation $h(\delta |\textbf{X})=0$ is guaranteed or not, and if it exists then how do we know that it is unique. This has been proved in the following theorem, and to the best of our knowledge, this hadn't been reported in the literature before. \\

\noindent  {\bf{Theorem 2.1}:} The solution to the equation $h(\delta |\textbf{X})=0$ exists and it is unique. 

\noindent {\bf{Proof}:} We start the proof by studying some basic behavior of the function $h(\delta |\textbf{X})$. 
\begin{itemize}
\item[(a)] Note that $\displaystyle{\lim_{\delta \to 0^{+}}} h(\delta |\textbf{X})= \infty$, since $(1/ \delta)\rightarrow \infty$, and  $\{ \textstyle \sum_1^n X_i^\delta (\textit{ln} X_i) / \sum_1^n X_i^\delta \}$ \\
$\rightarrow \{ \textstyle \sum_1^n \textit{ln} X_i/n \} = \textit{ln} \tilde{X}$. So, near $0$,  $h(\delta |\textbf{X})$ explodes to $\infty$. 
\item[(b)] Next, we study the behavior of $h(\delta |\textbf{X})$ as  $\delta\rightarrow \infty$. Suppose ($X_{(1)} \leq X_{(2)} \leq \dots \leq X_{(n)}$) be the order statistics of the observations. Note that the strict inequality (i.e.,  $X_{(1)} < X_{(2)} < \dots < X_{(n)}$) holds with probability 1. Therefore, 
\begin{equation}
\begin{split}
\displaystyle{\lim_{\delta \to \infty}} h(\delta |\textbf{X})& = 0 + \textit{ln} \tilde{X} - \displaystyle{\lim_{\delta \to \infty}}  \left[\{ \textstyle \sum_1^n X_{(i)}^\delta (\textit{ln} X_{(i)}) \} / \{ \textstyle \sum_1^n X_{(i)}^\delta \} \right] \\
& = \textit{ln} \tilde{X}-A(\textbf{X}) \: \text{(say)}, 
\end{split}
\end{equation}
where 
\begin{equation}
A(\textbf{X}) = \displaystyle{\lim_{\delta \to \infty}} \left[ \{ \textstyle \sum_{i=1}^{(n-1)} u_i^\delta (\textit{ln} X_{(i)}) + \textit{ln} X_{(n)} \} / \{ \textstyle \sum_{i=1}^{(n-1)} u_i^\delta + 1 \} \right]
\end{equation}
where $u_i = \left\lbrace X_{(i)} / X_{(n)} \right\rbrace \in (0,1)$ for $1 \leq i \leq (n-1)$. As $\delta \rightarrow \infty, u_i^\delta \rightarrow 0$, and hence 
\begin{equation}
A(\textbf{X}) = \textit{ln} X_{(n)}.
\end{equation}
Thus, 
\begin{equation}
\displaystyle{\lim_{\delta \to \infty}}  h(\delta |\textbf{X}) = \textit{ln} \tilde{X} - \textit{ln} X_{(n)} < 0.
\end{equation}
In other words, for $\delta$ large, $h(\delta |\textbf{X})$ takes a negative value.
\item[(c)] We now show that $h(\delta |\textbf{X})$ is decreasing in $\delta$.
\begin{equation} \label{eq:decrease}
h'(\delta |\textbf{X}) = -\delta^{-2} - [ (\textstyle \sum_1^n X_{i}^\delta ) { X_{i}^\delta (\textit{ln} X_i)^2 } - { \textstyle \sum_i^n  X_{i}^\delta (\textit{ln} X_i) }^2 ]/( \textstyle \sum_i^n X_{i}^\delta )^2
\end{equation}
\end{itemize}
Define $a_i=(\textit{ln} X_i) X_i^{\delta/2}$ and $b_i = X_i^{\delta/2}$ for $1 \leq i \leq n$. Thus, the expression inside$\ [ \ ]$ in the numerator of the second term in \eqref{eq:decrease} above is $\left( \textstyle \sum_i^n a_i^2 \right) \left( \textstyle \sum_i^n b_i^2 \right) - \left( \textstyle \sum_i^n a_i b_i \right)^2$, which is $\ge 0$, by Cauchy–Schwarz inequality. Thus $h'(\delta |\textbf{X}) \le 0$.

From the above (a) - (c) it is seen that $h(\delta |\textbf{X})$ takes a high positive value near 0, monotonically decreases as $\delta$ increase, and then takes negative values for large $\delta$. Therefore, there exists a unique $\delta$ value $(=\hat{\delta})$ for which $h(\delta|\textbf{X})=0$, and this completes the proof.
\section{Approximating the Sampling Distributions of $\hat{\delta}$ \& $(\hat{\beta}/\beta)$} 
\subsection{A Visual Representation of the Sampling Distributions}
While there are some asymptotic analytical as well as computational results on the bias and MSE of $\hat{\delta}$ and $\hat{\beta}$ as mentioned earlier, not much is known about their exact sampling distributions. This has motivated us to come up with a suitable approximation to the sampling distribution of $\hat{\delta}$ (and on the same token, that of $(\hat{\beta}/\beta)$ too) in a simple and convenient way. 

To start this investigation, we have fixed $\beta =1$, and have generated $\textbf{X}=(X_1, \dots, X_n)$ i.i.d from $W(\delta, \beta=1) $ a large number of times (says $M$ times). (W.l.g. $\beta=1$,  since the distribution of $\hat{\beta}$ with $\beta=1$ is the same as that of $(\hat{\beta}/\beta)$ for any $\beta$.) We have used $M=10^5$, $n=10(10)100$ and $\delta = 0.5(0.5) 10$. For a fixed $(n,\delta)$, and each of these $M$ replications, we have computed the value of $\hat{\delta}$ and $\hat{\beta}$, followed by their relative frequency histograms based on $M$ simulated values. Some of these plots have been presented in Figures 3.1 and 3.2.

\begin{figure}[h!]
	\includegraphics[width=15cm]{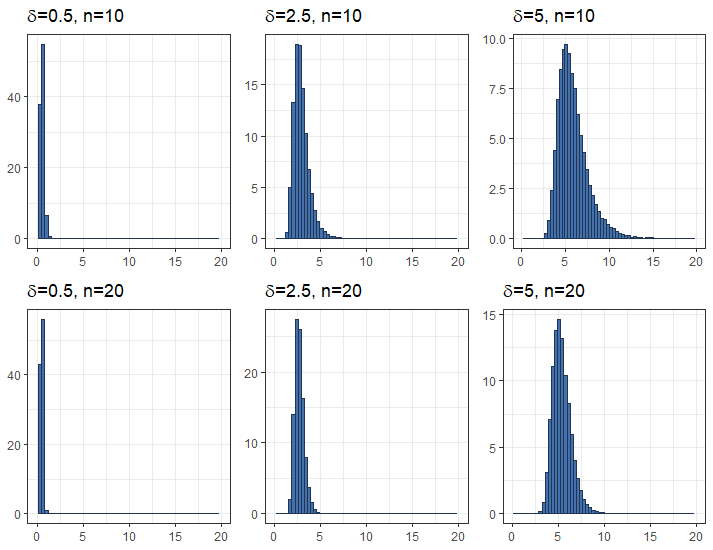}
	\centering
	\caption{Representative simulated relative frequency histograms of $\hat{\delta}$ ($10^5$ replications)}
	\label{fig:hist1}
\end{figure} 

\subsection{Approximating the sampling distribution of $\hat\delta$}

The relative frequency histograms of $\hat\delta$ in Figure \ref{fig:hist1} motivate us to hypothesize that the sampling distribution of $\hat\delta (=\hat{\delta}_{MLE})$ can be approximated fairy well by a suitable $W(a,b)$ where $a=a(n,\delta)$ and $b=b(n,\delta)$ are functions of $(n,\delta)$.

The following Tables 3.1 and 3.2 show some of the values of $a$ and $b$ for various $n$ and $\delta$ where a moment matching method as described below has been employed to approximate $a$ and $b$ based on $M=10^5$ replications for each $(n,\delta)$ combination.

For fixed  $(n,\delta)$, generate $\textbf{X}^{(m)}=(X_1^{(m)}, \dots,X_n^{(m)})$ iid from $W(\delta,1)$ in the $m^{th}$ replication, $1\leq m\leq M$. Based on $\textbf{X}^{(m)}$  obtain the MLEs of $\delta$ and $\beta$ as $\hat\delta^{(m)} (=\hat{\delta}^{(m)}_{MLE})$ and $\hat\beta^{(m)} (=\hat{\beta}^{(m)}_{MLE})$, respectively.

Now,  $\hat\delta^{(m)}$, $1\leq m\leq M$, are thought to be following $W(a,b)$ with mean $E(\hat\delta)\approx b\Gamma(1+1/a)$  and $Var(\hat\delta)\approx b^2 [\Gamma (1+2/a)-\{\Gamma(1+1/a)\}^2]$. Further, $E(\hat\delta)$ and $Var(\hat\delta)$ are approximated by $\bar{\hat{\delta}}^{(.)}=\sum_{m=1}^{M}\hat{\delta}^{(m)}/M$ and $s^2(\hat\delta)=\sum_{m=1}^{M}(\hat{\delta}^{(m)}-\bar{\hat{\delta}}^{(.)})^2/(M-1)$. Therefore, using the simulated $\hat{\delta}^{(m)}$, $1\leq m\leq M$, the parameters $a$ and $b$ are found by solving
\begin{equation}
	b\Gamma(1+1/a)=\bar{\hat{\delta}}^{(.)}\\
	\text{ and }
	b^2\left[\Gamma(1+2/a)-\{\Gamma(1+1/a)\}^2\right]=s^2(\hat{\delta}).
	\end{equation} \\
The following plots (Figures \ref{hist1} and \ref{hist2}) of  $a=a(n,\delta)$ and $b=b(n,\delta)$ against $n$ (for various $\delta$) and $\delta$ (for various $n$) show some interesting patterns. In all these plots $n$ varies from 10 to 100 with an increment of 10, and $\delta$ varies from 0.5 to 10.0 with an increment of 0.5.\\

\noindent {\bf{Remark 3.1}} Note that the tabulated values of $a(n,\delta)$ and $b(n,\delta)$ are subject to two sources of variations, the first one being the simulation variation while the observations were generated from $W(\delta,1)$, and the second one is the computational error incurred while solving the nonlinear equation $h(\delta|\textbf{X})=0$ (in \eqref{h_function}) numerically (however small this error might be). See the subsequent Subsection 3.4 on the magnitude of variations in the tabulated values in Tables 3.1 and 3.2 (as well as those in Tables 3.3 and 3.4 discussed in the next section).

\begin{figure}[H]
	\includegraphics[width=15cm]{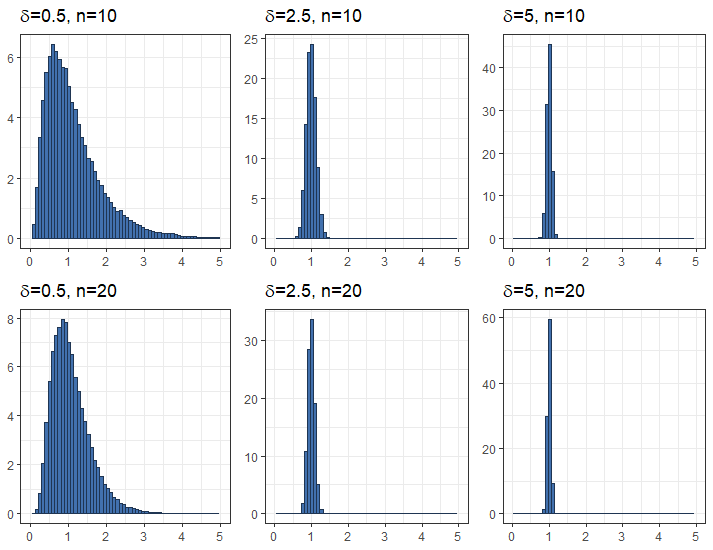}
	\centering
	\caption{Representative simulated relative frequency histograms of ($\hat{\beta}/\beta$) ($10^5$ replications)}
 \label{fig32}
\end{figure}

\begin{table}[H]\fontsize{10}{10}\selectfont
	\caption{Representative values of the shape parameter $a$ of $\hat{\delta } ~ \dot{\sim}~ W(a,b)$ using (3.1)}
	\label{tab1}
\begin{adjustbox}{max width=\textwidth}
	\centering
	\begin{tabular}{c|l|l|l|l|l|l|l|l|l|l|l|l}
		\hline
		$a(n,\delta$)& \multicolumn{12}{|c}{$\delta$}    \\ \hline
         $n$ & 0.5 & 1.0 & 1.5 & 2.0 & 3.0 &4.0 & 5.0 &6.0&7.0& 8.0&9.0 & 10.0  \\ \hline
        10 & 3.277 & 3.236 & 3.261 & 3.248 & 3.253 & 3.261 & 3.256 & 3.248 & 3.241 & 3.227 & 3.246 & 3.236 \\ 
        20 & 5.026 & 5.066 & 5.012 & 5.085 & 5.031 & 5.030 & 5.044 & 5.061 & 5.059 & 5.037 & 5.045 & 5.105 \\ 
        30 & 6.322 & 6.371 & 6.417 & 6.355 & 6.396 & 6.375 & 6.399 & 6.373 & 6.388 & 6.373 & 6.382 & 6.363 \\ 
        40 & 7.534 & 7.492 & 7.462 & 7.574 & 7.503 & 7.532 & 7.488 & 7.512 & 7.536 & 7.560 & 7.440 & 7.480 \\ 
        50 & 8.477 & 8.434 & 8.476 & 8.444 & 8.474 & 8.423 & 8.425 & 8.441 & 8.473 & 8.432 & 8.434 & 8.469 \\ 
        60 & 9.385 & 9.405 & 9.391 & 9.367 & 9.336 & 9.300 & 9.387 & 9.348 & 9.305 & 9.341 & 9.415 & 9.342 \\ 
        70 & 10.136 & 10.142 & 10.190 & 10.275 & 10.111 & 10.229 & 10.153 & 10.212 & 10.198 & 10.221 & 10.184 & 10.209 \\ 
        80 & 11.035 & 10.842 & 10.947 & 10.892 & 11.046 & 10.963 & 10.879 & 10.939 & 10.917 & 10.893 & 10.975 & 10.935 \\ 
        90 & 11.649 & 11.655 & 11.699 & 11.606 & 11.594 & 11.577 & 11.639 & 11.614 & 11.651 & 11.632 & 11.633 & 11.641 \\ 
        100 & 12.262 & 12.362 & 12.222 & 12.299 & 12.286 & 12.263 & 12.343 & 12.363 & 12.250 & 12.321 & 12.313 & 12.294 \\ \hline
	\end{tabular}
\end{adjustbox}

\end{table}

\begin{table}[H]\fontsize{10}{10}\selectfont
		\caption{Representative values of the scale parameter $b$ of $\hat{\delta}~\dot{\sim}~ W(a,b)$ using (3.1)}
		\label{tab2}
\begin{adjustbox}{max width=\textwidth}
	\centering
	\begin{tabular}{c|l|l|l|l|l|l|l|l|l|l|l|l}
		\hline
		$b(n,\delta$)& \multicolumn{12}{|c}{$\delta$}    \\ \hline
         $n$ & 0.5 & 1.0 & 1.5 & 2.0 & 3.0 &4.0 & 5.0 &6.0&7.0& 8.0&9.0 & 10.0  \\ \hline
       10 & 0.648 & 1.297 & 1.945 & 2.593 & 3.891 & 5.181 & 6.473 & 7.775 & 9.086 & 10.373 & 11.667 & 12.969 \\ 
        20 & 0.581 & 1.163 & 1.742 & 2.323 & 3.488 & 4.651 & 5.802 & 6.975 & 8.138 & 9.293 & 10.453 & 11.620 \\ 
        30 & 0.559 & 1.118 & 1.678 & 2.237 & 3.353 & 4.475 & 5.591 & 6.715 & 7.829 & 8.946 & 10.063 & 11.193 \\ 
        40 & 0.548 & 1.096 & 1.645 & 2.191 & 3.288 & 4.383 & 5.481 & 6.581 & 7.676 & 8.760 & 9.863 & 10.966 \\ 
        50 & 0.541 & 1.083 & 1.623 & 2.164 & 3.245 & 4.328 & 5.411 & 6.496 & 7.575 & 8.662 & 9.739 & 10.825 \\ 
        60 & 0.536 & 1.073 & 1.609 & 2.145 & 3.218 & 4.292 & 5.362 & 6.436 & 7.507 & 8.582 & 9.654 & 10.730 \\ 
        70 & 0.533 & 1.065 & 1.597 & 2.131 & 3.198 & 4.264 & 5.328 & 6.391 & 7.459 & 8.522 & 9.585 & 10.654 \\ 
        80 & 0.530 & 1.061 & 1.590 & 2.121 & 3.179 & 4.238 & 5.297 & 6.362 & 7.421 & 8.476 & 9.541 & 10.597 \\ 
        90 & 0.528 & 1.055 & 1.583 & 2.111 & 3.166 & 4.220 & 5.275 & 6.333 & 7.390 & 8.451 & 9.500 & 10.557 \\ 
        100 & 0.526 & 1.052 & 1.578 & 2.104 & 3.157 & 4.210 & 5.259 & 6.311 & 7.367 & 8.420 & 9.468 & 10.517 \\ \hline
	\end{tabular}
\end{adjustbox}
\end{table}



\begin{figure}[H]
 \begin{subfigure}{0.49\textwidth}
     \includegraphics[width=\textwidth]{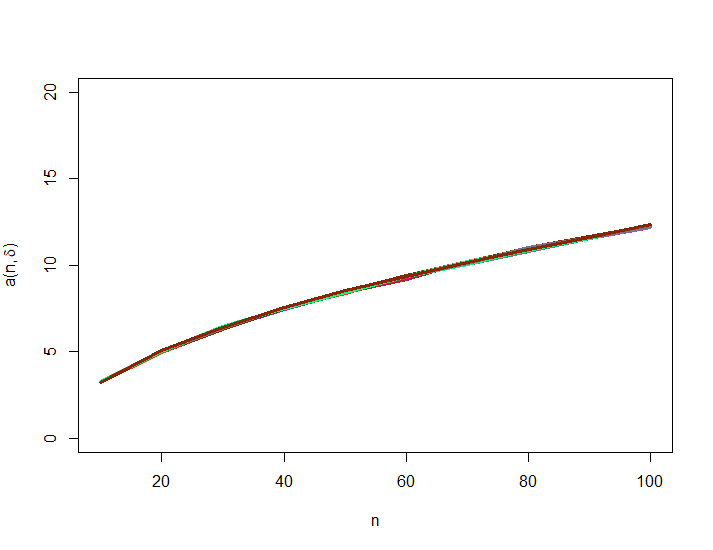}
     \caption{}
     \label{fig:hist4}
 \end{subfigure}
 \hfill
 \begin{subfigure}{0.49\textwidth}
     \includegraphics[width=\textwidth]{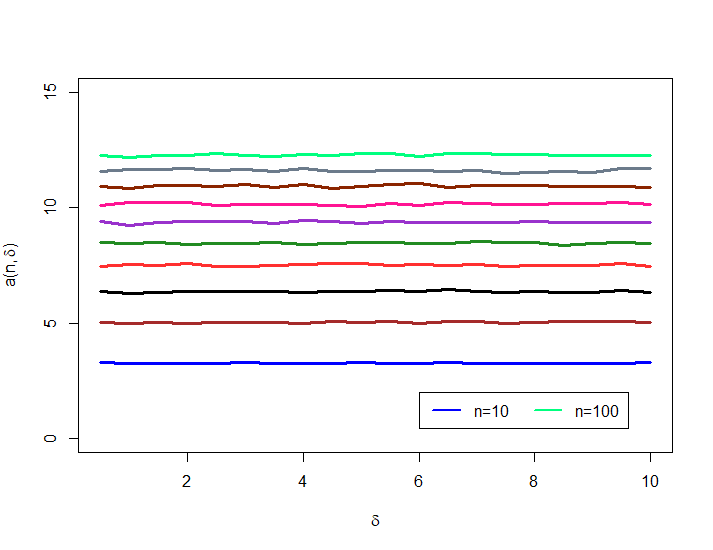}
     \caption{}
     \label{fig:hist3}
 \end{subfigure}
\caption{Plots of $a=a(n,\delta)$: (a) against  $n$ for fixed $\delta$: (b) against $\delta$ for fixed $n$}
 \label{hist1}
\end{figure}

Interestingly, $a=a(n,\delta)$ is pretty much constant against $\delta$ when $n$ is fixed, and it increases w.r.t $n$ in a sightly concave pattern for each fixed $\delta$. This motivates us to fit the following regression line (after several trial and error attempts) for $a=a(n,\delta)$ as
\begin{equation} \label{eq:rega}
{{a=a_0 + a_1 \textit{ln}(n+a_2)+\varepsilon,}}
\end{equation}
for suitable coefficients $a_0$, $a_1$ and $a_2$.

Based on the values of $a=a(n,\delta)$ as shown in Table \ref{tab1}, the above regression equation (\ref*{eq:rega}) gives a staggering R-square of almost $0.99$ indicating a very good fit. The following Figure \ref{res5}(\subref{res1}) shows the relative frequency histogram of the residuals from the regression model (\ref{eq:rega}).

\begin{figure}[H]
 \begin{subfigure}{0.49\textwidth}
     \includegraphics[width=\textwidth]{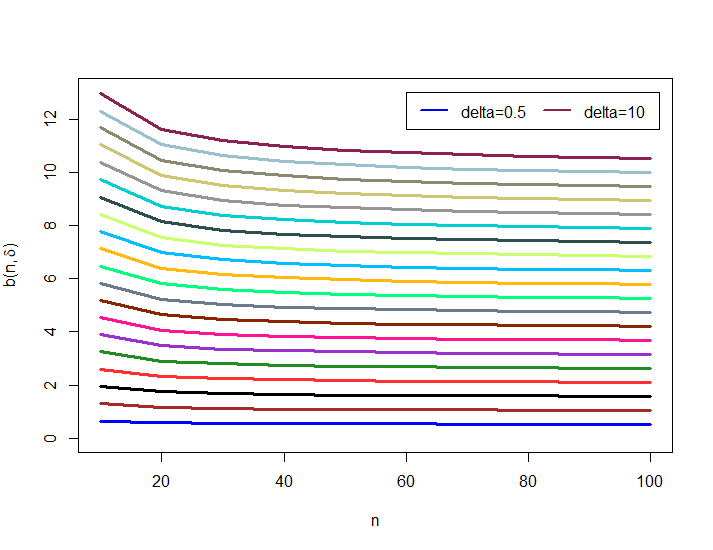}
     \caption{}
     \label{fig:hist7}
 \end{subfigure}
 \hfill
 \begin{subfigure}{0.49\textwidth}
     \includegraphics[width=\textwidth]{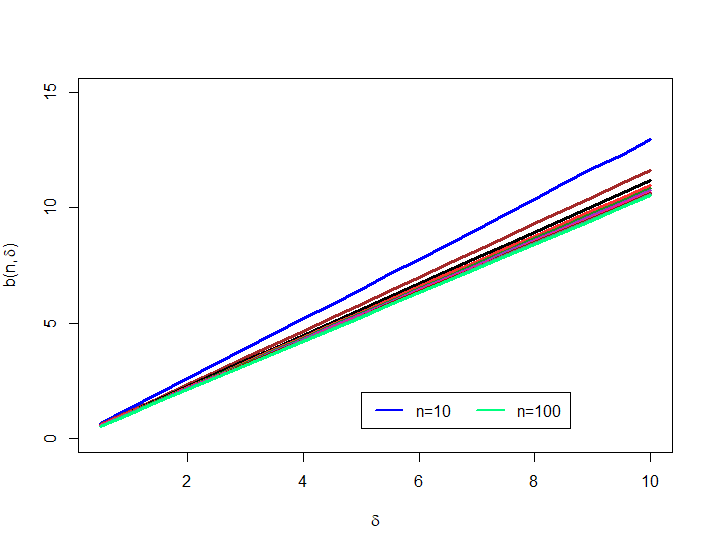}
     \caption{}
     \label{fig:hist5}
 \end{subfigure}
\caption{Plots of $b=b(n,\delta)$: (a) against  $n$ for fixed $\delta$: (b) against $\delta$ for fixed $n$}
 \label{hist2}
\end{figure}

Similar to the simulated values of $a$, we now proceed to approximate $b=b(n,\delta)$ with a suitable regression equation. Figures \ref{hist2} (\subref{fig:hist7}) – (\subref{fig:hist5}) show the concave pattern of $b$ against $n$ for fixed $\delta$, and the straight line pattern of $b$ against $\delta$ for fixed $n$. Unlike $a$, $b$ seems to be mildly decreasing w.r.t $n$; however, it seems parallel (as a function of $n$) for two different values of $\delta$. On the other hand, as a function of $\delta$, $b$ appears to be strictly linearly increasing with different slopes for different $n$. The following regression plane for $b=b(n,\delta)$ has been proposed:
\begin{equation} \label{eq:regb}
	b=b_0  \delta  / \{1+b_1 \textit{ln}(\textit{ln}(n)) \}+\varepsilon,
\end{equation}
for suitable coefficients $b_0$ and $b_1$.

Based on the values of $b=b(n,\delta)$ as shown in Table \ref{tab2}, the above regression equation (\ref*{eq:regb}) again gives a staggering R-square of almost $0.99$ indicating a very good fit. The relative frequency histogram of the residuals from the regression model (\ref*{eq:regb}) is given in Figure \ref{res5}(\subref{res2}).
The approximated coefficients, free from $n$, used in (\ref*{eq:rega}) and (\ref*{eq:regb}) have been provided in Table \ref{tab5}.
\subsection{Approximating the sampling distribution of ($\hat\beta/\beta$)}

Based on the histograms of ($\hat\beta/\beta$) in Figure \ref{fig32} we hypothesize that $(\hat\beta/\beta) ~\dot{\sim}~ W(c,d)$. The following Tables \ref{tab3} and \ref{tab4} show the values of $c$ and $d$ for various $n$ and $\delta$, where a moment matching method, as described earlier, has been employed to approximate $c$ and $d$ based on $M=10^5$ replications for each $(n,\delta)$ combination.

\begin{table}[H]\fontsize{10}{10}\selectfont
	\caption{Representative values of the shape parameter $c$ of $(\hat\beta/\beta) ~ \dot{\sim}~ W(c,d)$}
	\label{tab3}
\begin{adjustbox}{max width=\textwidth}
	\centering
	\begin{tabular}{c|l|l|l|l|l|l|l|l|l|l|l|l}
		\hline
		$c(n,\delta$)& \multicolumn{12}{|c}{$\delta$}    \\ \hline
        $n$& 0.5 & 1.0 & 1.5 & 2.0 & 3.0 & 4.0 & 5.0 & 6.0 & 7.0 & 8.0 & 9.0 & 10.0 \\ \hline
        10 & 1.589 & 3.179 & 4.761 & 6.396 & 9.510 & 12.707 & 15.979 & 19.168 & 22.171 & 25.603 & 28.663 & 31.749 \\ 
        20 & 2.227 & 4.445 & 6.717 & 8.880 & 13.303 & 17.815 & 22.315 & 26.572 & 31.021 & 35.815 & 39.820 & 44.726 \\ 
        30 & 2.718 & 5.429 & 8.243 & 10.875 & 16.182 & 21.810 & 27.113 & 32.574 & 37.905 & 43.272 & 48.835 & 54.069 \\
        40 & 3.099 & 6.223 & 9.288 & 12.446 & 18.777 & 24.976 & 31.275 & 37.138 & 43.464 & 49.875 & 55.598 & 62.192 \\ 
        50 & 3.465 & 6.958 & 10.403 & 13.805 & 20.786 & 27.787 & 34.368 & 41.722 & 48.365 & 55.051 & 62.463 & 69.671 \\ 
        60 & 3.761 & 7.551 & 11.343 & 15.143 & 22.612 & 30.413 & 37.551 & 45.586 & 52.849 & 60.678 & 68.425 & 76.392 \\
        70 & 4.094 & 8.162 & 12.172 & 16.298 & 24.288 & 32.828 & 40.683 & 48.924 & 57.224 & 65.549 & 73.558 & 81.722 \\ 
        80 & 4.356 & 8.708 & 13.123 & 17.624 & 26.087 & 34.735 & 43.346 & 52.215 & 60.911 & 69.820 & 78.170 & 87.402 \\ 
        90 & 4.624 & 9.290 & 13.838 & 18.367 & 27.547 & 36.881 & 46.040 & 55.844 & 64.295 & 74.029 & 83.129 & 92.288 \\ 
        100 & 4.863 & 9.712 & 14.661 & 19.239 & 29.235 & 39.112 & 48.656 & 58.387 & 68.075 & 77.528 & 87.944 & 96.944 \\ \hline
	\end{tabular}
\end{adjustbox}
\end{table}

Interestingly, $c=c(n,\delta)$ is pretty much linear against $\delta$ when $n$ is fixed, and it increases w.r.t $n$ in a sightly concave pattern for each fixed $\delta$. This motivates us to fit the following regression line (after several trial and error attempts) for $c=c(n,\delta)$ as
\begin{equation} \label{eq:regc}
{{c=c_0  \delta \textit{ln}(n+c_1)+\varepsilon,}}
\end{equation}
for suitable coefficients $c_0$ and $c_1$.

Based on the values of $c=c(n,\delta)$ in Table \ref{tab3}, the above regression equation (\ref*{eq:regc}) gives a staggering R-square of almost $0.98$, again indicating a very good fit. The following Figure \ref{res5}(\subref{res3}) shows the relative frequency histogram of the residuals from the regression model (\ref{eq:regc}). \\

Similar to the simulated values of $c$, we now proceed to approximate $d=d(n,\delta)$ with a suitable regression equation. Figures \ref{hist4} (\subref{hist3.7}) – (\subref{hist3.8}) show that $d$ is mildly decreasing w.r.t. both $n$ and $\delta$, but soon stabilizes at 1. The following regression plane for $d=d(n,\delta)$ has been proposed:
\begin{equation} \label{eq:regd}
    d= d_0 + d_1/n + d_2/\delta + d_3/(\delta n) +\varepsilon,
\end{equation}
for suitable coefficients $d_0$, $d_1$, $d_2$ and $d_3$.

Based on the values of $d=d(n,\delta)$ as shown in Table \ref{tab4}, the above regression equation (\ref*{eq:regd}) gives again staggering R-square value of $0.99$ indicating a very good fit. The relative frequency histogram of the residuals from the regression model (\ref*{eq:regd}) is given in Figure \ref{res5}(\subref{res4}).

Again, the approximated values of the coefficients used in (\ref*{eq:regc}) and (\ref*{eq:regd}) have been provided in Table \ref{tab5}, and these are free from $n$.

\begin{table}[H]\fontsize{10}{10}\selectfont
		\caption{Representative values of the scale parameter $d$ of $(\hat\beta/\beta)~\dot{\sim}~ W(c,d)$}
		\label{tab4}
\begin{adjustbox}{max width=\textwidth}
	\centering
	\begin{tabular}{c|l|l|l|l|l|l|l|l|l|l|l|l}
		\hline
		$d(n,\delta$)& \multicolumn{12}{|c}{$\delta$}    \\ \hline
        $n$ &0.5 & 1.0 & 1.5 & 2.0 & 3.0 & 4.0 & 5.0 & 6.0 & 7.0 & 8.0 & 9.0 & 10.0 \\ \hline
        10 & 1.284 & 1.132 & 1.088 & 1.064 & 1.044 & 1.032 & 1.025 & 1.021 & 1.018 & 1.016 & 1.014 & 1.013 \\ 
        20 & 1.218 & 1.103 & 1.066 & 1.050 & 1.033 & 1.025 & 1.020 & 1.017 & 1.014 & 1.012 & 1.011 & 1.010 \\ 
        30 & 1.178 & 1.086 & 1.056 & 1.043 & 1.027 & 1.021 & 1.017 & 1.014 & 1.012 & 1.010 & 1.009 & 1.008 \\ 
        40 & 1.163 & 1.076 & 1.050 & 1.037 & 1.025 & 1.018 & 1.015 & 1.012 & 1.011 & 1.009 & 1.008 & 1.007 \\ 
        50 & 1.142 & 1.067 & 1.046 & 1.033 & 1.022 & 1.017 & 1.013 & 1.011 & 1.009 & 1.008 & 1.008 & 1.007 \\ 
        60 & 1.128 & 1.062 & 1.042 & 1.031 & 1.021 & 1.015 & 1.013 & 1.010 & 1.009 & 1.008 & 1.007 & 1.006 \\ 
        70 & 1.123 & 1.058 & 1.039 & 1.029 & 1.019 & 1.014 & 1.012 & 1.010 & 1.008 & 1.007 & 1.006 & 1.006 \\ 
        80 & 1.112 & 1.055 & 1.036 & 1.027 & 1.018 & 1.014 & 1.011 & 1.009 & 1.008 & 1.007 & 1.006 & 1.005 \\ 
        90 & 1.107 & 1.052 & 1.035 & 1.026 & 1.017 & 1.013 & 1.010 & 1.009 & 1.007 & 1.006 & 1.006 & 1.005 \\ 
        100 & 1.102 & 1.050 & 1.033 & 1.025 & 1.016 & 1.012 & 1.010 & 1.008 & 1.007 & 1.006 & 1.005 & 1.005 \\ \hline
	\end{tabular}
\end{adjustbox}
\end{table}

\begin{figure}[H]
 \begin{subfigure}{0.49\textwidth}
     \includegraphics[width=\textwidth]{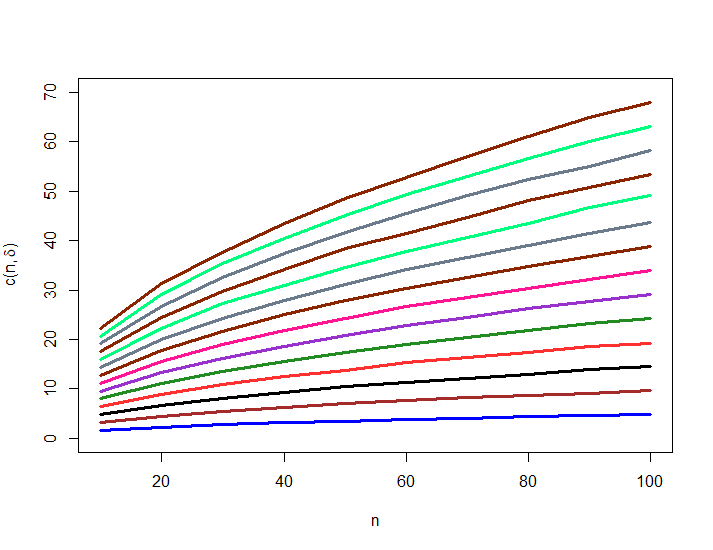}
     \caption{}
     \label{hist3.5}
 \end{subfigure}
 \hfill
 \begin{subfigure}{0.49\textwidth}
     \includegraphics[width=\textwidth]{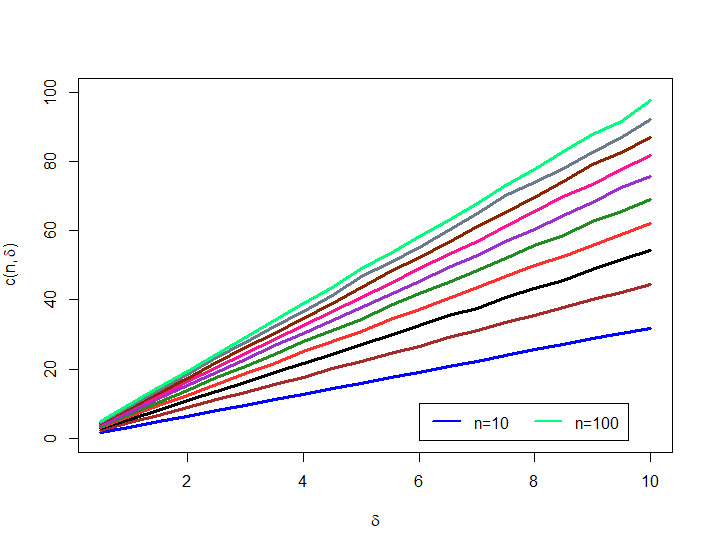}
     \caption{}
     \label{hist3.6}
 \end{subfigure}
\caption{Plots of $c=c(n,\delta)$: (a) against  $n$ for fixed $\delta$: (b) against $\delta$ for fixed $n$}
 \label{hist3}
\end{figure}

To ascertain how well our four proposed regression models (3.2 – 3.5) fit $a=a(n,\delta)$, $b=b(n,\delta)$, $d=d(n,\delta)$ and $c=c(n,\delta)$, we have carried out the standard normality tests on the residuals for each regression model. Note that we have a total 200 residuals based on the combinations of $n$ and $\delta$. The following Table 3.5 presents the observed $R^2$ as well as the p-values based on the Shapiro-Wilk test (SWT) and the Anderson–Darling test (ADT) to test normality.\\

It appears that our proposed model (3.5) for $d=d(n,\delta)$ is not unbiased even though it gives an overall good fit. This may be due to the fact that (3.5) doesn't capture the curvature of $d=d(n,\delta)$, for small values of $n$, as a function of $\delta$.

\begin{figure}[H]
 \begin{subfigure}{0.49\textwidth}
     \includegraphics[width=\textwidth]{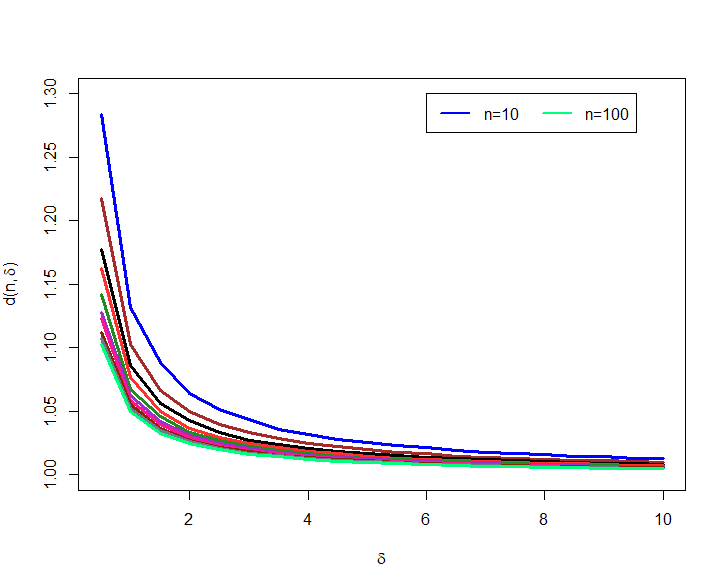}
     \caption{}
     \label{hist3.7}
 \end{subfigure}
 \hfill
 \begin{subfigure}{0.49\textwidth}
     \includegraphics[width=\textwidth]{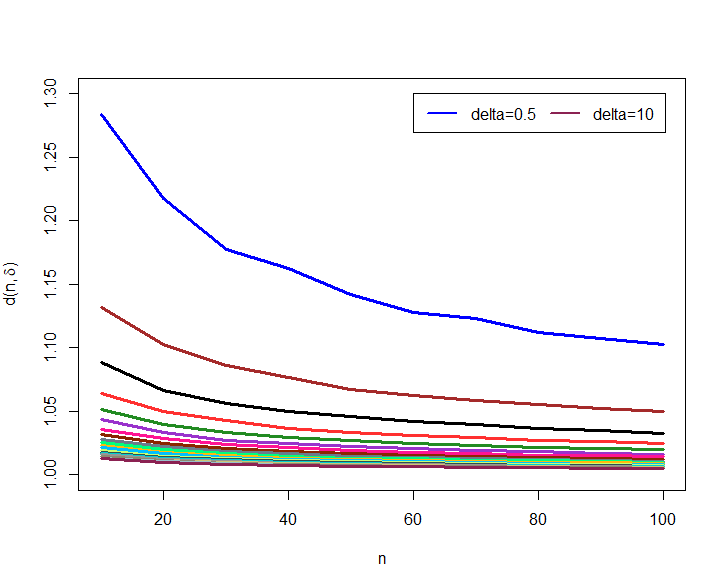}
     \caption{}
     \label{hist3.8}
 \end{subfigure}
\caption{Plots of $d=d(n,\delta)$: (a) against  $n$ for fixed $\delta$: (b) against $\delta$ for fixed $n$}
 \label{hist4}
\end{figure}

\begin{figure}[H]
 \begin{subfigure}{0.49\textwidth}
     \includegraphics[width=\textwidth]{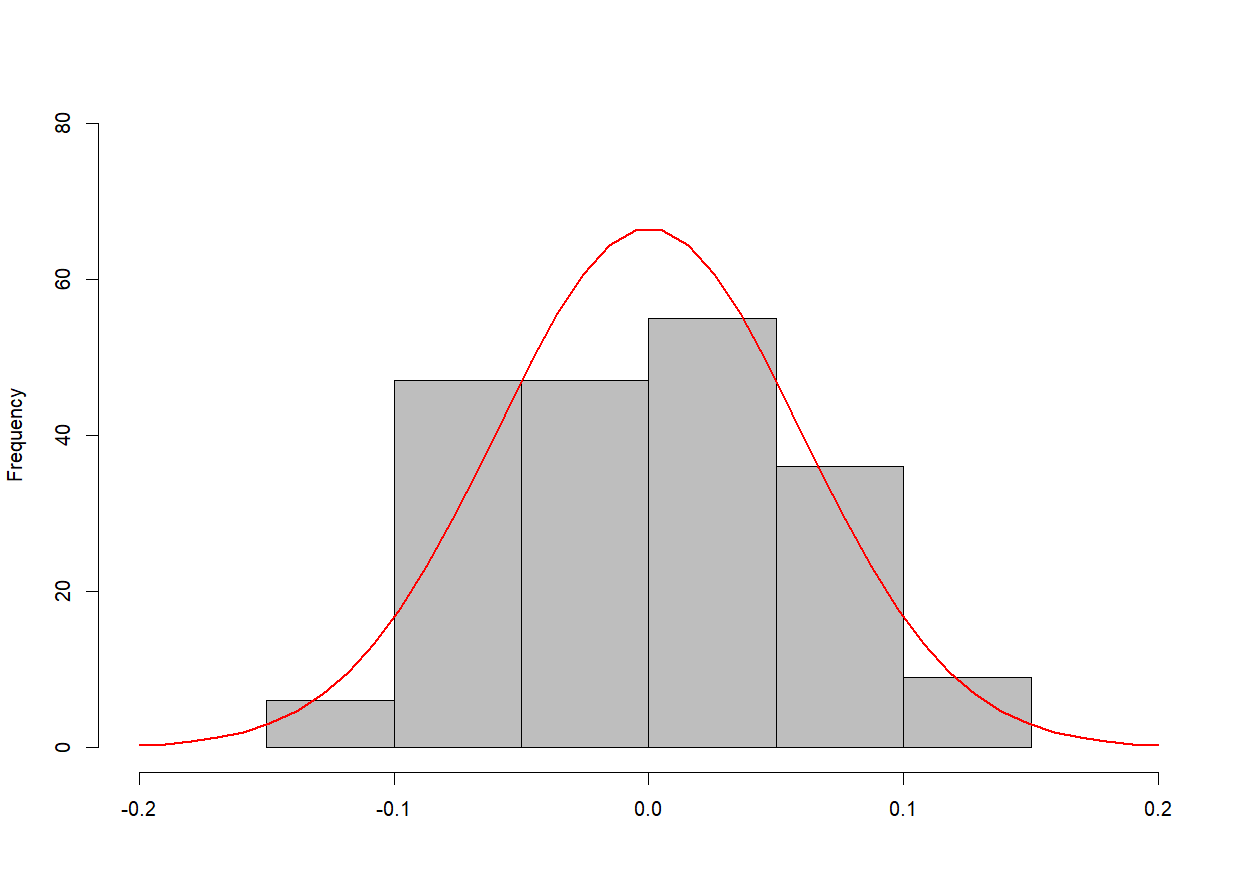}
     \caption{}
     \label{res1}
 \end{subfigure}
 \hfill
 \begin{subfigure}{0.49\textwidth}
     \includegraphics[width=\textwidth]{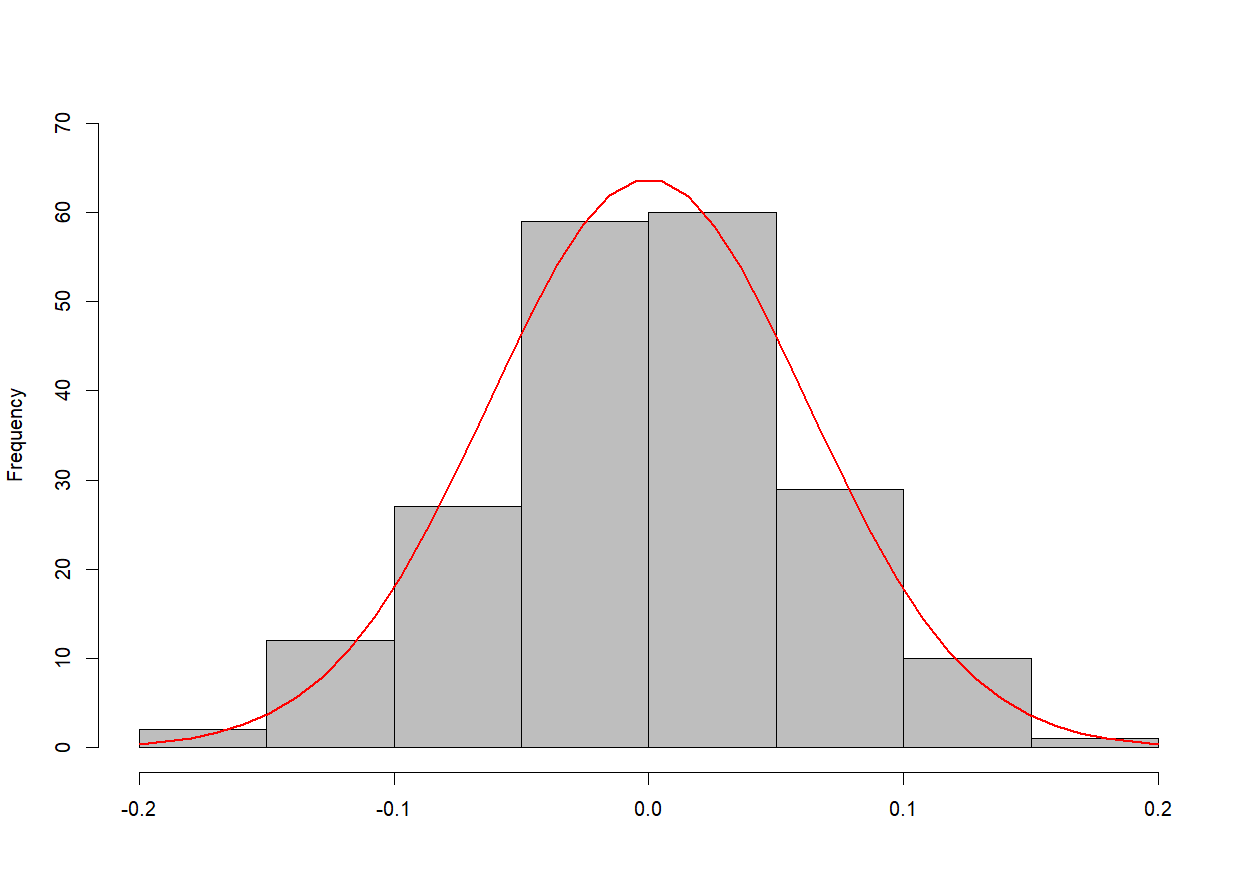}
     \caption{}
     \label{res2}
 \end{subfigure}
 
 \medskip
 \begin{subfigure}{0.49\textwidth}
     \includegraphics[width=\textwidth]{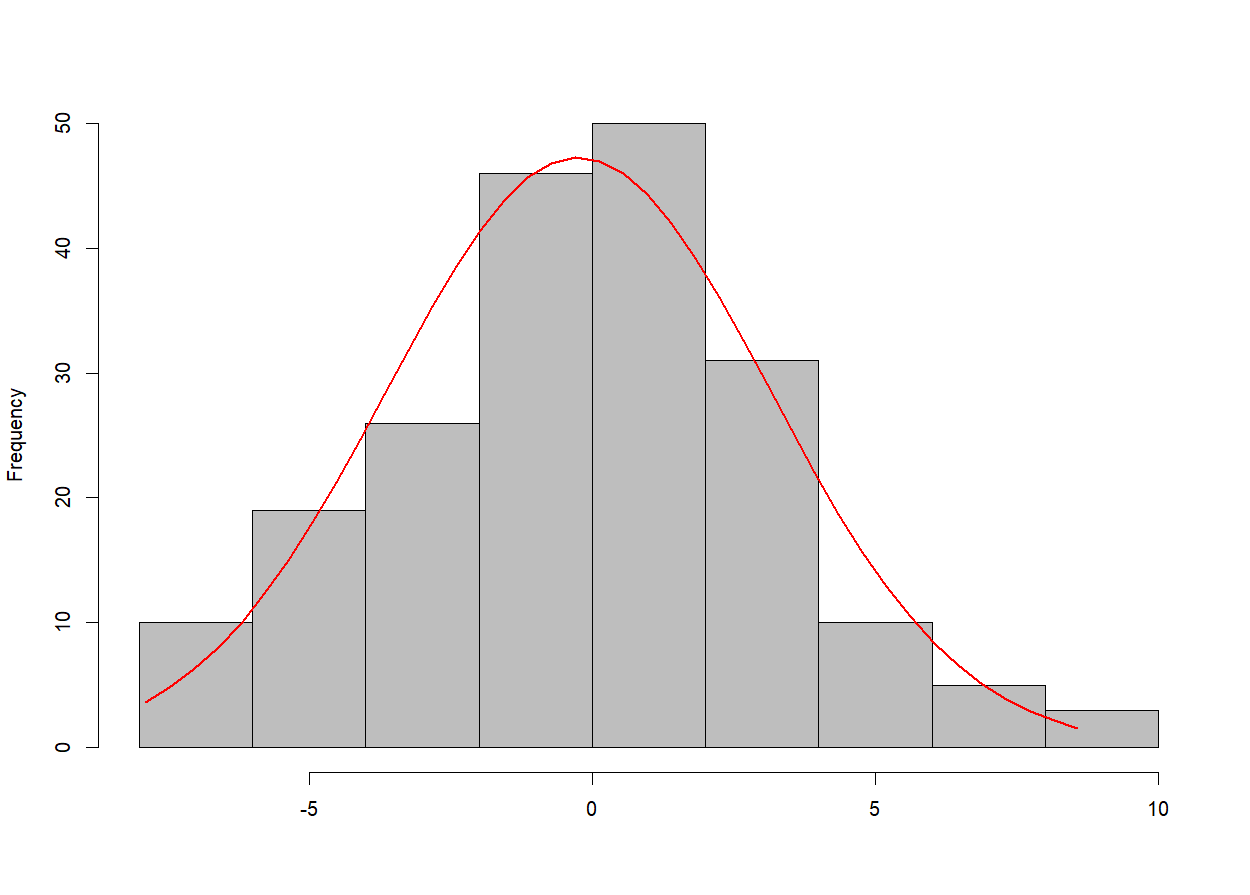}
     \caption{}
     \label{res3}
 \end{subfigure}
 \hfill
 \begin{subfigure}{0.49\textwidth}
     \includegraphics[width=\textwidth]{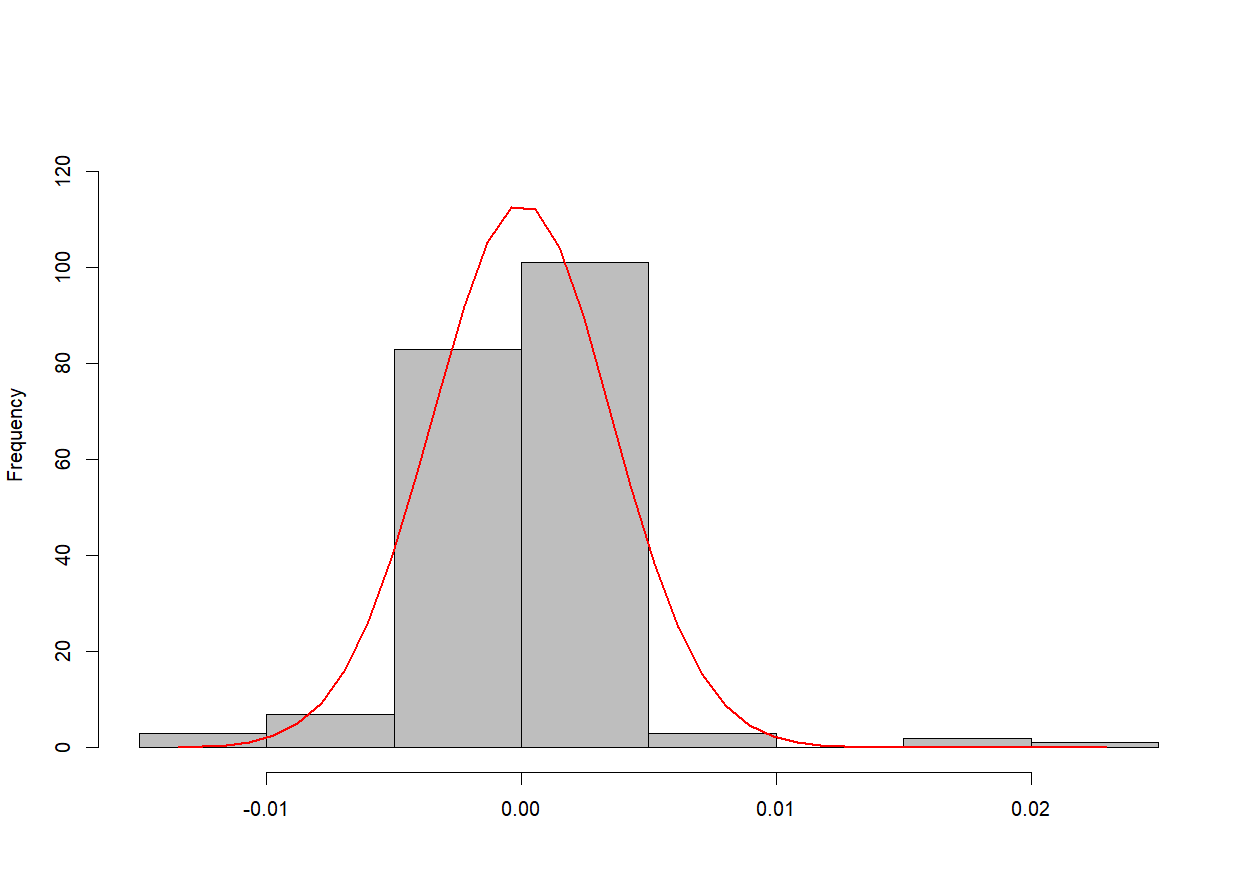}
     \caption{}
     \label{res4}
 \end{subfigure}
\caption{Relative frequency histogram of the residuals (a) (\ref{eq:rega}); (b) (\ref{eq:regb}); (c) (\ref{eq:regc}); (d) (\ref{eq:regd}).}
 \label{res5}

\end{figure}

\begin{table}[H]\fontsize{10}{10}\selectfont
    \centering
	\begin{tabular}{l|l|c|c|c|c}
		\hline
		Model & For & Estimated Coefficients  & Observed $R^2$ & SWT p-value & ADT p-value    \\  \hline
        (3.2) & $a(n,\delta)$& $\hat{a}_0=-26.485$ &  0.99 & 0.497&  0.325  \\
         & & $\hat{a}_1=7.915$&  & &    \\
         & &$\hat{a}_2=33.125$ &  & &   \\  \hline
        (3.3) & $b(n,\delta)$&$\hat{b}_0=1.775$ & 0.99 & 0.290 & 0.208  \\  
         & & $\hat{b}_1=0.463$&  & &    \\  \hline
        (3.4) & $c(n,\delta)$&$\hat{c}_0= 1.944$ &  0.98  & 0.325 & 0.507\\
         & &$\hat{c}_1=-5.782$ &  & &   \\  \hline
        (3.5) & $d(n,\delta)$&$\hat{d}_0 = 0.999$ & 0.99 & $<$0.001 & $<$0.001  \\ 
        & &$\hat{d}_1 = -0.031$ &  & &   \\
        & & $\hat{d}_2= 0.048$&  & &    \\
        & &$\hat{d}_3= 1.003$ &  & &   \\ \hline
	\end{tabular}
 	\caption{An evaluation summary of our proposed regression models }
	\label{tab5}
\end{table}

\subsection{Simulation Variations of the Sampling Distribution Parameters}
In the earlier subsections, we have provided the values of $a=a(n, \delta)$, $b=b(n, \delta)$, $c=c(n, \delta)$ and $d=d(n, \delta)$ which are the parameters of the sampling distributions of $\hat{\delta}$ and $(\hat{\beta}/\beta)$ based on the hypothesis that $\hat{\delta} ~ \dot{\sim}~ W(a, b)$ and $(\hat{\beta}/\beta) ~ \dot{\sim}~ W(c, d)$, respectively. The tabulated values of $a, b, c$, and $d$ as shown in Tables 3.1 – 3.4 are based on $10^5$ replications each. However. these tabulated values can vary from one run of $10^5$ replications to another, and hence one might be interested in assessing the magnitude of simulation variations of $a, b, c$, and $d$. An extended repeated simulation with $100$ runs for each combination of $n$ and $\delta$ (based on $10^5$ replications) was carried out, and some representative results are provided in Table 3.6. The tabulated values are the mean, the standard deviation (SD) as well as the coefficient of variation (CV).\\

\begin{table}[H]\fontsize{10}{10}\selectfont

\begin{adjustbox}{max width=\textwidth}
	\centering
    \centering
    \begin{tabular}{c|c|c|c|c| c|c|c|c|c}
    \hline
        $n$& $\delta$ &$a(n,\delta)$ &  & $b(n,\delta)$ & & $c(n,\delta)$& & $d(n,\delta)$ \\ \hline
        & & $mean (SD)$ & $CV$ & $mean (SD)$ &$CV$ & $mean (SD)$ &$CV$ & $mean (SD)$& $CV$\\ \hline
        10 & 0.5 & 3.2492 (0.0215) & 0.0066 &0.6480 (0.0006) &0.0009 & 1.5942 (0.0047) &0.0029 & 1.2860 (0.0026)& 0.0020 \\ 
         & 5.0 & 3.2457 (0.0205) &0.0063 &6.4800 (0.0069) &0.0011 & 15.9421 (0.0439)&0.0028 & 1.0254 (0.0002) & 0.0002 \\ 
         & 10.0 & 3.2453 (0.0220) &0.0068 &12.9564 (0.0135) & 0.0010& 33.4905 (0.0818) & 0.0024& 1.0126 (0.0001) &0.0001 \\ \hline
        50 & 0.5 & 8.4754 (0.0309) & 0.0036 &0.5411 (0.0002) & 0.0004& 3.4649 (0.0115) & 0.0033& 1.1420 (0.0011) &0.0010 \\ 
         & 5.0 & 8.4816 (0.0358) &0.0042 &5.4113 (0.0020) & 0.0004& 34.6509 (0.1090) &0.0031 & 1.0133 (0.0001) & 0.0001 \\ 
        & 10.0 & 8.4819 (0.0333) &0.0039 & 11.3638 (0.0044) &0.0004 & 72.7450 (0.2111) & 0.0029& 1.0063 (0.0001) &0.0001 \\ \hline
        100 & 0.5 & 12.2927 (0.0488) &0.0040 &  0.5260 (0.0001) &0.0003 & 4.8601 (0.0149) & 0.0031& 1.1020 (0.0008) & 0.0007\\ 
         & 5.0 & 12.2877 (0.0489) & 0.0040 & 5.2601 (0.0014) & 0.0003& 48.5900 (0.1556) & 0.0032& 1.0100 (0.0001) & 0.0001 \\ 
         & 10.0 & 12.2868 (0.0435) & 0.0035& 10.5199 (0.0027) & 0.0003&97.2039 (0.3392) &0.0035 &1.0049 (0.0000) & 0.0000\\ \hline
	\end{tabular}
\end{adjustbox}
	\caption{Mean, standard deviation (SD) and CV (= $SD/mean$) based on 100 runs (of the simulation with $10^5$ replications)}
	\label{tab3.6}
\end{table} 
\newpage
When we read values in Tables 3.1 – 3.4, they are pretty much within one SD of the mean as seen in Table 3.6. For example, with $n=10$ and $\delta=0.5$, the value of $a(n, \delta)$, i.e., 3.277 is about one SD $(= 0.0215)$ away from the mean $(= 3.2492)$. The same holds true for $b, c$, and $d$.\\
It might seem that the SD is increasing with $n$, but this is a bit misleading as some of the sampling distribution parameters (especially, $a, b$, and $c$) are also increasing with $n$. But when we look at the relative magnitude of the variation, as measured in terms of CV, it is seen to be decreasing generally as $n$ increases. 

\section{An Evaluation of the Goodness of the Sampling Distributions of $\hat\delta$ \& $(\hat\beta/\beta)$}
In the previous section, we have discussed approximating the sampling distributions of $\hat\delta$ and $(\hat\beta/\beta)$ as
\begin{equation}\label{eq4.1}
	\hat{\delta}~\dot{\sim}~ W(a,b) \text{ and } (\hat\beta/\beta) ~\dot{\sim}~ W(c,d),
\end{equation}
where Tables 3.1 and 3.2 provide the simulated values of ($a$, $b$) and ($c$, $d$) based on $M=10^5$ replications. In other words, for each combination of $(n,\delta)$, $M=10^5$ replications (each of size $n$) were used from $W(\delta,\beta=1)$, and then $(\hat\delta,\hat\beta)$ values were computed for each sample. These $M=10^5$ values of $(\hat\delta,\hat\beta)$ (i.e, $(\hat\delta^{(m)},\hat\beta^{(m)}), 1\leq m\leq M$) were used to obtain $(a,b)$ and $(c,d)$ through the moment matching technique. In this section, we are going to use those replicated $(\hat{\delta}^{(m)}, \hat{\beta}^{(m)})$, $1\leq m \leq M$, values for each combination of $(\delta, \beta)$ to investigate how well the hypothesized Weibull sampling distributions of $\hat{\delta}$ and $(\hat{\beta}/ \beta )$ work. Note that, in subsections 3.2 and 3.3 we have provided further approximations of $(a,b)$ and $(c,d)$ based on regression models, respectively.

\subsection{Goodness of $W(a, b)$ as the Approximate Distribution of $\hat\delta$}
It is customary to compare two distributions by their percentile values. If $\hat{\delta}~\dot{\sim}~ W(a,b)$, then the $(100p)^{th}$ percentile $(0<p<1)$ of $W(a,b)$ is 
\begin{equation}\label{eq4.2}
	b\{-\textit{ln}(1-p)\}^{1/a}=Q_p(\hat{\delta}|a,b)= \hat Q_p \text{ \quad (say)  },
\end{equation}
where $a=a(n,\delta)$ and $b=b(n,\delta)$ have been listed in Tables 3.1 and 3.2. We now compare the above  $\hat Q_p $ value with the empirical $(100p)^{th}$ percentile of $\hat\delta$ based on $\hat{\delta}^{(m)}$, $1\leq m \leq M=10^5$, defined as
\begin{equation} \label{eq4.3}
	\tilde{Q}_p=\tilde{Q}_p(\hat\delta)=(Mp)^{th} \text{ ordered value of } \{\hat\delta^{(m)}, 1 \leq m\leq M\}.
\end{equation}
Further, the regression equations \eqref{eq:rega} and \eqref{eq:regb} provide approximations of $a=a(n,\delta)$ and $b=b(n,\delta)$ as
\begin{equation} \label{eq4.4}
	a \approx \hat a(n,\delta)= \hat{a}_0 + \hat{a}_1  \textit{ln}(n+\hat{a}_2),
\end{equation} 
\begin{equation} \label{eq4.5}
	b \approx \hat b(n,\delta)= \hat{b}_0  \delta  / \{1+\hat{b}_1 \textit{ln}(\textit{ln}(n)) \}.
\end{equation} 
Therefore, $\hat{Q}_p$ in \eqref{eq4.2} can be approximated further as
\begin{equation}\label{eq4.6}
	\hat b\{-\textit{ln}(1-p)\}^{1/\hat a}=\hat {\hat {Q}}_p \text{ \quad (say)  },
\end{equation}
where $\hat a$ and $\hat b$ are given in \eqref{eq4.4} -- \eqref{eq4.5}. The following Table 4.1 compares the values of $\hat{Q}_p, ~\tilde{Q}_p$ and $\hat {\hat {Q}}_p$ for selected values of $(n, \delta)$ and $p~(=0.1, ~ 0.25, ~ 0.50, ~ 0.75 \text{ and } 0.90)$.  \\

The following three figures (Figure 4.1 – 4.3) show the representative plots comparing $\hat{Q}_p, ~\tilde{Q}_p$ and $\hat {\hat {Q}}_p$ as functions of $p$ (for fixed $(n,\delta)$), $\delta$ (for fixed $(n,p)$) and $n$ (for fixed $(n,\delta)$). For convenience $\hat{Q}_p, ~\tilde{Q}_p$, ~$\hat {\hat {Q}}_p$ have been identified as A, B, and C, respectively. \\

\noindent {\bf{Remark 4.1}}. Table 4.1, and corroborated by the representative plots of Figure 4.1 – 4.3, it is noted that (i)  $\hat{Q}_p$ and $\hat {\hat {Q}}_p$ are nearly identical irrespective of the values of $n, \delta$ and $p$. This shows a very good fit for the regression equations (3.2) and (3.3) to $a$ and $b$, i.e., the parameters of the hypothesized sampling distributions $W(a,b)$ for $\hat{\delta}$. (ii) Also, in terms of approximating $~\tilde{Q}_p$, $\hat{Q}_p$ and/or $\hat {\hat {Q}}_p$ are/is fairly close except for very small $p$ $(p<0.25)$, however such approximations get
better as expected when $n$ increases.

\subsection{Goodness of $W(c, d)$ as the Approximate Distribution of $(\hat\beta/\beta)$}
Similar to what has been done for $\hat{\delta}$, we are going to study the percentiles of $(\hat\beta/\beta)$. If ($\hat\beta/\beta)~\dot{\sim}~ W(c,d)$, then the $(100p)^{th}$ percentile $(0<p<1)$ of $W(c,d)$ is 
\begin{equation}\label{eq4.22}
	d\{-\textit{ln}(1-p)\}^{1/c}=\hat Q^*_p \text{ \quad (say)  },
\end{equation}
where $c=c(n,\delta)$ and $d=d(n,\delta)$ have been listed in Tables \ref{tab3} and \ref{tab4}. We now compare the above  $\hat Q^*_p $ value with the empirical $(100p)^{th}$ percentile of $(\hat\beta/\beta)$ based on, $1\leq m \leq M=10^5$, defined as
\begin{equation} \label{eq4.33}
	\tilde{Q}^*_p=\tilde{Q}^*_p(\hat\beta/\beta)=(Mp)^{th} \text{ ordered value of } \{(\hat\beta^{(m)}/\beta), 1 \leq m\leq M\}.
\end{equation}
Further, the regression equations \eqref{eq:regc} and \eqref{eq:regd} provide approximations of $c=c(n,\delta)$ and $d=d(n,\delta)$ as
\begin{equation} \label{eq4.44}
	c \approx \hat c(n,\delta)= \hat{c}_0  \delta  \textit{ln}(n+ \hat{c}_1),
\end{equation} 
\begin{equation} \label{eq4.55}
	d \approx \hat d(n,\delta)= \hat{d}_0+ \hat{d}_1/n+ \hat{d}_2/(\delta)+ \hat{d}_3/(\delta n).
\end{equation} 
Therefore, $\hat{Q}^*_p$ in \eqref{eq4.22} can be approximated further as
\begin{equation}\label{eq4.66}
	\hat d\{-\textit{ln}(1-p)\}^{1/\hat c}=\hat {\hat {Q}}^*_p \text{ \quad (say)  },
\end{equation}
where $\hat c$ and $\hat d$ are given in \eqref{eq4.44} – \eqref{eq4.55}. The following Table 4.2 compares the values of $\hat{Q}^*_p, ~\tilde{Q}^*_p$ and $\hat {\hat {Q}}^*_p$ for selected values of $(n, \delta)$ and $p~(=0.1, ~ 0.25, ~ 0.50, ~ 0.75\text{ and } 0.90)$.\\

The following three figures (Figure 4.4 – 4.6) show the representative plots comparing $\hat{Q}^*_p, ~\tilde{Q}^*_p$ and $\hat {\hat {Q}}^*_p$ as functions of $p$ (for fixed $(n,\delta)$), $\delta$ (for fixed $(n,p)$) and $n$ (for fixed $(n,\delta)$). \\

\noindent {\bf{Remark 4.2}}. Table 4.2, and as seen from the representative plots of Figure 4.4 - 4.6, it appears that the sampling distribution of $(\hat \beta/ \beta)$ gets approximated very well by our proposed $W(c,d)$ distribution, where one can use $c=c(n, \delta)$ and $d=d(n, \delta)$ either as listed in Tables 3.3 – 3.4, or the regression equations (3.4) – (3.5) irrespective of the values of $n, \delta$ and $p$.

\subsection{Further comparisons from First Order Point of View}

In this subsection, we are going to compare the approximate sampling distributions of $\hat{\delta}$ and $\hat{\beta}$ (or ($\hat{\beta}/\beta$)) as hypothesized in Section 3 with the asymptotic distributions of $\hat{\delta}$ and $\hat{\beta}$, respectively, in terms of \lq First Order Bias\rq (FOB) and \lq First Order Variance\rq (FOV).

Note that in (\ref{B1}) and (\ref{E_beta}) we had provided the FOB expressions of $\hat{\delta}$ and ($\hat{\beta}/\beta$), based on their exact sampling distributions (for large $n$) as 
\begin{equation}
    \text{FOB}(\hat{\delta})=(B_1/n) \text{~~ and~~ } \text{FOB}(\hat{\beta}/\beta)=(B_2/n), 
\end{equation}
where
\begin{equation}
    B_1=\delta(1.37953) \text{~~ and~~ } B_2=(1/\delta^2)(0.55433)-(1/\delta)(0.369815).
\end{equation}

Also, using the standard asymptotic theory, it is well known that FOV expressions of $\delta$ and ($\hat{\beta}/\beta$) (which are also commonly known  as the asymptotic variances, and are essentially the Cram\'{e}r-Rao lower bounds of the respective unbiased parameter estimates) as  
\begin{equation}
    \text{FOV}(\hat{\delta})=(V_1/n) \text{~~ and~~ } \text{FOV}(\hat{\beta}/\beta)=(V_2/n), 
\end{equation}
where
\begin{equation}
    V_1=\{\delta^2/\zeta(2)\} \text{~~ and~~ } V_2=\{A_1^2+\zeta(2)\}/\{\delta^2\zeta(2)\},
\end{equation}
where
    $A_1=1-\gamma,\gamma=0.577216\dots$ is the Euler's constant, and $\zeta(2)=\sum_{j=1}^{\infty}1/j^2=\pi^2/6$  is the Riemann zeta function with argument 2 (see Tanaka et al. (2018)).

\newpage

\begin{table}[H]
		\caption{Goodness of $W(a, b)$ as the Approximate Distribution of $\hat{\delta}$}.
	\label{table4.1}
 \begin{table}[H]\fontsize{9}{10}\selectfont
    \centering
    \rotatebox{90}{
   \begin{adjustbox}{max width=1.31\textwidth}
    \centering
    \begin{tabular}{|l|l|l|l|l|l|l|l|l|l|l|}
    \hline
         & &  \multicolumn{8}{|c|}{$\delta$} \\ 
         & &  \multicolumn{8}{|c|}{($\hat{Q}_p, ~\tilde{Q}_p$, ~$\hat {\hat {Q}}_p$)} \\ \hline
         $n$ & $p$ & 1.00 &2.00 & 3.00 & 6.00 & 7.00& 8.00 & 9.00& 10.00 \\ \hline
        10 & 0.10  & (2.08, 2.73, 2.09) & (2.20, 2.79, 2.20) &(2.15, 2.76, 2.12) & (2.20, 2.80, 2.20) &(2.65, 3.34, 2.64) &(2.21, 2.80, 2.23) & (2.84, 3.63, 2.87)& (2.27, 2.87, 2.27)  \\ 
        ~ & 0.25  & (2.97, 3.86, 3.03) & (3.11, 4.01, 3.08)&(3.09, 4.01, 3.05) & (3.11, 4.01, 3.08) &(3.63, 4.64, 3.57) &(3.12, 4.02, 3.13) & (3.98, 4.80, 3.95) & (3.13, 4.06, 3.19) \\ 
        ~ & 0.50  & (4.17, 5.09, 4.18) &(4.30, 5.17, 4.24)& (4.22, 5.11, 4.22) & (4.30, 5.17, 4.24) &(5.00, 5.70, 4.95) &(4.46, 5.22, 4.39) & (5.18, 6.02, 5.19)& (4.44, 5.34, 4.44) \\ 
        ~ & 0.75  & (5.53, 6.29, 5.51) &(5.75, 6.36, 5.71)& (5.73, 6.35, 5.65) & (5.75, 6.36, 5.71) &(6.47, 7.02, 6.49)&(5.78, 6.42, 5.73) &(7.05, 7.27, 7.03)& (5.83, 6.44, 5.77) \\ 
        ~ & 0.90  & (7.52, 7.62, 7.41) &(7.56, 7.89, 7.47)& (7.53, 7.73, 7.45) & (7.56, 7.89, 7.47) &(8.83, 8.84, 8.79) &(7.88, 8.02, 7.77) & (9.31, 9.33, 9.17)& (7.95, 8.02, 7.91) \\ \hline
        20 & 0.10  & (2.27, 2.62, 2.29) &(2.36, 2.78, 2.40)& (2.33, 2.69, 2.33) & (2.36, 2.78, 2.40) &(2.91, 3.36, 2.94)&(2.48, 2.79, 2.51) &(3.14, 3.57, 3.20)& (2.56, 2.85, 2.56) \\ 
        ~ & 0.25  & (3.28, 3.73, 3.28) &(3.43, 3.93, 3.48)& (3.35, 3.78, 3.36) & (3.43, 3.93, 3.48) &(3.94, 4.36, 3.96)& (3.48, 3.96, 3.49) &(4.12, 4.71, 4.18)& (3.54, 4.04, 3.56) \\ 
        ~ & 0.50  & (4.34, 4.10, 4.40) & (4.42, 5.08, 4.42)& (4.34, 5.05, 4.41) & (4.42, 5.08, 4.42) &(4.96, 5.59, 5.03)&(4.47, 5.08, 4.49) &(5.22, 5.89, 5.23)& (4.52, 5.12, 4.53) \\ 
        ~ & 0.75  & (5.50, 6.04, 5.57) &(5.56, 6.06, 5.61)& (5.52, 6.05, 5.60) & (5.56, 6.06, 5.61) &(6.30, 6.73, 6.36)& (5.58, 6.30, 5.70) &(6.49, 6.98, 6.56)& (5.79, 6.30, 5.82) \\ 
        ~ & 0.90  & (6.82, 7.15, 6.91) &(7.02, 7.35, 7.10)& (6.85, 7.24, 6.96) & (7.02, 7.35, 7.10) &(7.85, 7.98, 8.01)& (7.05, 7.40, 7.10) &(8.17, 8.37, 8.24)& (7.07, 7.44, 7.21) \\ \hline
        30 & 0.10  & (2.36, 2.59, 2.39) &(2.47, 2.67, 2.48)& (2.46, 2.64, 2.46) & (2.47, 2.67, 2.48) &(2.10, 3.29, 3.01)& (2.55, 2.81, 2.59) &(3.19, 3.45, 3.23)& (2.64, 2.86, 2.67) \\ 
        ~ & 0.25  & (3.52, 3.75, 3.54) &(3.53, 3.77, 3.57)& (3.53, 3.76, 3.56) & (3.53, 3.77, 3.57) &(4.08, 4.50, 4.11)& (3.54, 3.88, 3.58) &(4.23, 4.65, 4.28)& (3.55, 3.99, 3.61) \\ 
        ~ & 0.50  & (4.45, 4.89, 4.52) &(4.60, 5.03, 4.65)& (4.56, 4.93, 4.60) & (4.60, 5.03, 4.65) &(5.10, 5.62, 5.16)& (4.60, 5.14, 4.65) &(5.28, 5.71, 5.35)& (4.64, 5.15, 4.72) \\ 
        ~ & 0.75  & (5.50, 6.02, 5.56) &(5.61, 6.10, 5.67)& (5.52, 6.04, 5.58) & (5.61, 6.10, 5.67) &(6.32, 6.72, 6.38)& (5.69, 6.17, 5.77) &(6.44, 6.89, 6.51)&(5.73, 6.20, 5.82) \\ 
        ~ & 0.90  & (6.68, 7.14, 6.75) &(6.90, 7.32, 6.98)& (6.87, 7.23, 6.95) & (6.90, 7.32, 6.98) &(7.65, 7.80, 7.75)&(7.01, 7.34, 7.08) &(7.86, 8.19, 7.94)&(7.02, 7.39, 7.11) \\ \hline
        40 & 0.10  & (2.44, 2.56, 2.46) &(2.53, 2.73, 2.56)& (2.45, 2.63, 2.47) & (2.53, 2.73, 2.56) &(3.06, 3.30, 3.09)& (2.59, 2.79, 2.60) &(3.25, 3.51, 3.28)& (2.61, 2.81, 2.63)  \\
        ~ & 0.25  & (3.44, 3.77, 3.47) &(3.65, 3.82, 3.67)& (3.57, 3.78, 3.59) & (3.65, 3.82, 3.69) &(4.06, 4.38, 4.10)& (3.65, 3.88, 3.69) &(4.28, 4.63, 4.33)& (3.68, 3.91, 3.71) \\ 
        ~ & 0.50  & (4.47, 4.84, 4.51) & (4.64, 5.03, 4.68)&(4.58, 4.91, 4.62) & (4.64, 5.03, 4.68) &(5.16, 5.59, 5.20)&(4.69, 5.05, 4.74) &(5.32, 5.70, 5.39)& (4.70, 5.06, 4.75) \\ 
        ~ & 0.75  & (5.56, 6.06, 5.62) & (5.70, 6.14, 5.73)& (5.66, 6.12, 5.69) & (5.70, 6.14, 5.73) &(6.27, 6.65, 6.32)& (5.72, 6.15, 5.76) &(6.50, 6.99, 6.55)& (5.73, 6.15, 5.78) \\ 
        ~ & 0.90  & (6.74, 7.06, 6.80) & (6.87, 7.17, 6.93)& (6.79, 7.14, 6.85) & (6.87, 7.17, 6.93) &(7.43, 7.77, 7.49)& (6.91, 7.26, 6.96) &(7.83, 7.98, 7.90)& (6.97, 7.27,  7.03) \\ \hline
        50 & 0.10  & (2.47, 2.56, 2.48) &(2.51, 2.70, 2.53)& (2.49, 2.66, 2.51) & (2.51, 2.70, 2.53) &(3.08, 3.30, 3.09)& (2.59, 2.75, 2.60) &(3.27, 3.54, 3.29)& (2.59, 2.79, 2.61) \\ 
        ~ & 0.25  & (3.46, 3.71, 3.48) &(3.63, 3.84, 3.65)& (3.58, 3.78, 3.60) & (3.63, 3.84, 3.65) &(4.14, 4.38, 4.17)& (3.69, 3.85, 3.71) &(4.31 ,4.70, 4.33)& (3.73, 3.99, 3.76) \\ 
        ~ & 0.50  & (4.57, 4.78, 4.59) &(4.67, 4.95, 4.69)& (4.57, 4.88, 4.60) & (4.67, 4.95, 4.69) &(5.18, 5.55, 5.21)& (4.67, 4.96, 4.70) &(5.38, 5.70, 5.40)& (4.72, 5.09, 4.75) \\ 
        ~ & 0.75  & (5.62, 5.98, 5.65) &(5.70, 6.07, 5.73)& (5.65, 6.06, 5.70) & (5.70, 6.07, 5.73) &(6.22, 6.62, 6.25)& (5.72, 6.11, 5.75)&(6.49, 6.81, 6.52)& (5.80, 6.14, 5.84) \\ 
        ~ & 0.90  & (6.73, 7.10, 6.77) &(6.78, 7.15, 6.81)& (6.75, 7.14, 6.78) & (6.77, 7.15, 6.81) &(7.46, 7.69, 7.51)&(6.85, 7.17, 6.90) &(7.78, 8.04, 7.81)& (7.01, 7.25, 7.05) \\ \hline
    \end{tabular}
   \end{adjustbox}{}}
\end{table}
\end{table}

\begin{table}[H]
\ContinuedFloat
		\caption{ (cont.): Goodness of $W(a, b)$ as the Approximate Distribution of $\hat{\delta}$}
 \begin{table}[H]\fontsize{9}{10}\selectfont
    \centering
    \rotatebox{90}{
   \begin{adjustbox}{max width=1.31\textwidth}
 \centering
    \begin{tabular}{|l|l|l|l|l|l|l|l|l|l|l|}
    \hline
         & &  \multicolumn{8}{|c|}{$\delta$} \\ 
         & &  \multicolumn{8}{|c|}{($\hat{Q}_p, ~\tilde{Q}_p$, ~$\hat {\hat {Q}}_p$)} \\ \hline
         $n$ & $p$& 1.00 &2.00 & 3.00 & 6.00 & 7.00& 8.00 & 9.00& 10.00 \\ \hline
       60 & 0.10  & (2.41, 2.60, 2.41) &(2.53, 2.69, 2.54)& (2.55, 2.62, 2.53) & (2.53, 2.69, 2.54) &(3.09, 3.27, 3.10)& (2.58, 2.73, 2.58) &(3.29, 3.49, 3.29)&(2.62, 2.84, 2.63)   \\ 
        ~ & 0.25  & (3.52, 3.65, 3.52) &(3.61, 3.81, 3.62)& (3.61, 3.80, 3.62) & (3.61, 3.81, 3.62) &(4.22, 4.36, 4.22)&(3.62, 3.89, 3.62) &(4.30, 4.58, 4.31)&(3.75, 3.93, 3.76) \\ 
        ~ & 0.50  & (4.64, 4.85, 4.65) &(4.68, 4.90, 4.69)& (4.64, 4.86, 4.65) & (4.68, 4.90, 4.69) &(5.16, 5.45, 5.17)&(4.70, 4.92, 4.70) &(5.46, 5.70, 5.48)& (4.70, 5.08, 4.70) \\ 
        ~ & 0.75  & (5.63, 5.99, 5.64) &(5.68, 6.05, 5.69)& (5.67, 6.04, 5.68) & (5.68, 6.05, 5.69) &(6.25, 6.60, 6.26)& (5.86, 6.08, 5.87) &(6.56, 6.89, 6.58)&(5.86, 6.10, 5.87) \\ 
        ~ & 0.90  & (6.67, 7.11, 6.67) &(6.75, 7.16, 6.76)& (6.71, 7.11, 6.71) & (6.75, 7.16, 6.76) &(7.51, 7.70, 7.53)& (6.94, 7.20, 6.95) &(7.74, 8.05, 7.75)& (7.04, 7.26, 7.04) \\ \hline
        70 &  0.10  & (2.38, 2.56, 2.38) &(2.56, 2.70, 2.56)& (2.47, 2.62, 2.47) & (2.56, 2.70, 2.56) &(3.08, 3.25, 3.08)& (2.57, 2.70, 2.56) &(3.30, 3.49, 3.29)& (2.72, 2.86, 2.71) \\ 
        ~ & 0.25  & (3.56, 3.70, 3.55)&(3.60, 3.81, 3.59)& (3.57, 3.78, 3.58) & (3.60, 3.81, 3.59) &(4.16, 4.36, 4.14)& (3.72, 3.84, 3.70) &(4.37, 4.56, 4.37)& (3.72, 3.93, 3.73) \\ 
        ~ & 0.50  & (4.63, 4.80, 4.61) &(4.71, 4.94, 4.70)& (4.70, 4.87, 4.69) & (4.71, 4.94, 4.70) &(5.18, 5.40, 5.17)& (4.75, 4.96, 4.73) &(5.50, 5.74, 5.49)& (4.75, 5.07, 4.74) \\ 
        ~ & 0.75  & (5.65, 5.95, 5.64) &(5.75, 6.08, 5.76)& (5.66, 5.96, 5.65) & (5.75, 6.08, 5.76) &(6.37, 6.57, 6.34)& (5.78, 6.11, 5.77) &(6.53, 6.80, 6.51)& (5.94, 6.17, 5.92) \\ 
        ~ & 0.90  & (6.76, 7.04, 6.75) &(6.82, 7.14, 6.79)& (6.79, 7.09, 6.77) & (6.82, 7.14, 6.79) &(7.51, 7.67, 7.50)&(6.84, 7.15, 6.82) &(7.70, 8.04, 7.68)& (6.93, 7.21, 6.92) \\ \hline
        80 & 0.10  & (2.43, 2.53, 2.42) &(2.56, 2.69, 2.55)& (2.44, 2.64, 2.42) & (2.56, 2.69, 2.55) &(3.08, 3.22, 3.06)& (2.59, 2.72, 2.57) &(3.31, 3.52, 3.29)& (2.73, 2.84, 2.71) \\ 
        ~ & 0.25  & (3.51, 3.74, 3.50) &(3.64, 3.76, 3.62)& (3.59, 3.76, 3.57) & (3.64, 3.76, 3.62) &(4.18, 4.39, 4.15)& (3.65, 3.81, 3.64) &(4.37, 4.55, 4.34)& (3.78, 3.95, 3.76) \\ 
        ~ & 0.50  & (4.61, 4.84, 4.59)&(4.72, 4.10, 4.70)& (4.68, 4.84, 4.66) & (4.72, 4.99, 4.70) &(5.21, 5.44, 5.19)&(4.74, 4.99, 4.72) &(5.49, 5.72, 5.45)&(4.85, 5.02, 4.82) \\ 
        ~ & 0.75  & (5.65, 5.89, 5.62) &(5.72, 6.07, 5.69)& (5.68, 5.91, 5.64) & (5.72, 6.07, 5.69) &(6.29, 6.60, 6.26)&(5.86, 6.16, 5.83) &(6.55, 6.82, 6.52)& (5.92, 6.20, 5.88) \\ 
        ~ & 0.90  & (6.71, 7.06, 6.66) &(6.87, 7.08, 6.83)&(6.87, 7.08, 6.82) & (6.88, 7.08, 6.83) &(7.44, 7.64, 7.39)& (6.89, 7.15, 6.86) &(7.66, 7.96, 7.63)& (6.96, 7.25, 6.92) \\ \hline
        90 &  0.10  & (2.45, 2.53, 2.43) &(2.56, 2.67, 2.53)&(2.50, 2.66, 2.46) & (2.56, 2.67, 2.53) &(3.07, 3.21, 3.04)& (2.61, 2.73, 2.59) &(3.32, 3.49, 3.29)&(2.72, 2.82, 2.69) \\ 
        ~ & 0.25  & (3.55, 3.72, 3.52) &(3.61, 3.78, 3.58)&(3.58, 3.74, 3.55) & (3.61, 3.78, 3.58) &(4.21, 4.32, 4.18)&(3.69, 3.82, 3.65) &(4.35, 4.56, 4.31)&(3.79, 3.95, 3.76) \\ 
        ~ & 0.50  & (4.62, 4.81, 4.58) &(4.74, 4.95, 4.70)&(4.64, 4.88, 4.60) & (4.74, 4.95, 4.70) &(5.22, 5.47, 5.18)& (4.79, 5.02, 4.74) &(5.43, 5.73, 5.38)& (4.82, 5.04, 4.77) \\ 
        ~ & 0.75  & (5.69, 5.88, 5.63)&(5.73, 6.06, 5.67)& (5.69, 5.92, 5.64) & (5.73, 6.06, 5.67) &(6.38, 6.64, 6.32)&(5.80, 6.19, 5.75) &(6.55, 6.82, 6.48)& (5.97, 6.20, 5.92) \\
        ~ & 0.90  & (6.75, 7.08, 6.69) &(6.95, 7.15, 6.89)&(6.81, 7.09, 6.75) & (6.95, 7.15, 6.89) &(7.40, 7.65, 7.33)&(6.96, 7.17, 6.90) &(7.67, 7.98, 7.60)&(6.96, 7.29, 6.90) \\ \hline
       100 &  0.10  & (2.48, 2.54, 2.45) &(2.55, 2.67, 2.52)&(2.70, 2.80, 2.67) & (2.98, 3.12, 2.95) &(3.06, 3.20, 3.03)&(3.25, 3.36, 3.21) &(3.33, 3.46, 3.38)&(3.45, 3.56, 3.412) \\ 
        ~ & 0.25  & (3.57, 3.67, 3.53) &(3.62, 3.81, 3.57)& (3.78, 3.92, 3.74) & (4.09, 4.26, 4.04) &(4.17, 4.31, 4.13)& (4.32, 4.48, 4.27) &(4.38, 4.58, 4.32)& (4.55, 4.66, 4.49) \\ 
        ~ & 0.50  & (4.60, 4.79, 4.55) &(4.75, 4.90, 4.69)& (4.82, 5.04, 4.76) & (5.18, 5.35, 5.12) &(5.26, 5.50, 5.19)&(5.37, 5.55, 5.30) &(5.42, 5.75, 5.35)& (5.63, 5.79, 5.57) \\ 
        ~ & 0.75  & (5.71, 5.94, 5.63)&(5.78, 6.05, 5.71)& (5.94, 6.16, 5.87) & (6.19, 6.41, 6.12) &(6.32, 6.68, 6.25)& (6.52, 6.71, 6.42) &(6.57, 6.78, 6.49)& (6.64, 6.92, 6.56) \\ 
        ~ & 0.90  & (6.75, 7.07, 6.68) &(6.90, 7.18, 6.82)&(7.02, 7.27, 6.93) & (7.32, 7.57, 7.23) &(7.44, 7.66, 7.35)& (7.57, 7.84, 7.48) &(7.66, 7.99, 7.57)& (7.85, 8.04, 7.75) \\ \hline
    \end{tabular}
   \end{adjustbox}{}}
\end{table}
\end{table}

\begin{figure}[H]
    \includegraphics[width=15cm]{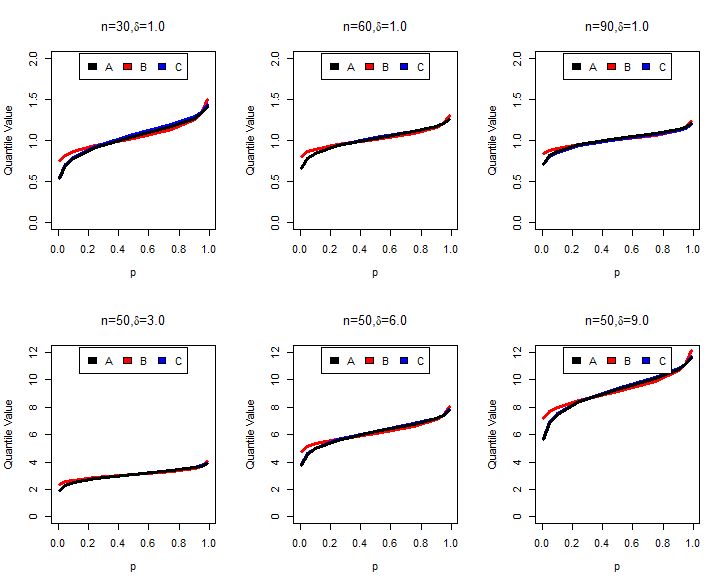}
    \centering
	\caption{Plots of $\hat{Q}_p$(= A)$, ~\tilde{Q}_p$(= B) and $\hat {\hat {Q}}_p(= C)$ as functions of $p$}
 
    \includegraphics[width=15cm]{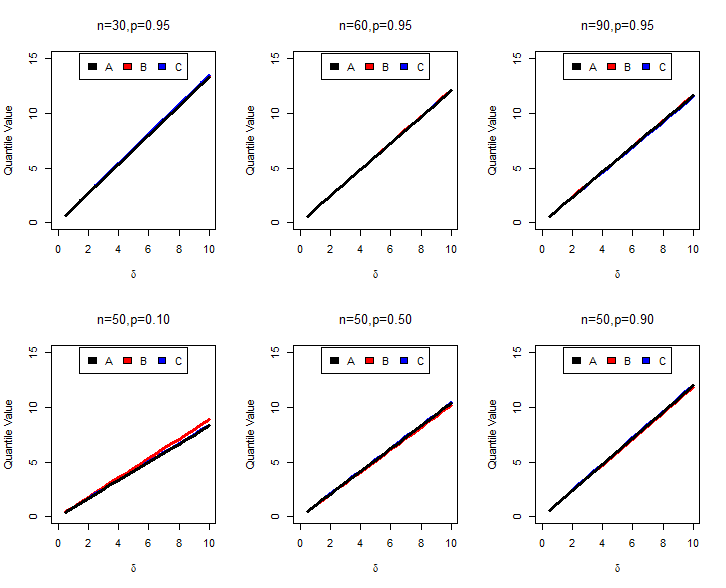}
    \centering
	\caption{Plots of $\hat{Q}_p$(= A)$, ~\tilde{Q}_p$(= B) and $\hat {\hat {Q}}_p(= C)$  as functions of $\delta$}
\end{figure}

\newpage

\begin{table}[H]
		\caption{Goodness of $W(c, d)$ as the Approximate Distribution of $(\hat{\beta}/\beta)$}
	\label{table4.2}
 \begin{table}[H]\fontsize{9}{10}\selectfont
    \centering
    \rotatebox{90}{
   \begin{adjustbox}{max width=1.31\textwidth}
 \centering
    \begin{tabular}{|l|l|l|l|l|l|l|l|l|l|l|}
    \hline
         & &  \multicolumn{8}{|c|}{$\delta$} \\ 
         & &  \multicolumn{8}{|c|}{($\hat{Q}^*_p, ~\tilde{Q}^*_p$, ~$\hat {\hat {Q}}^*_p$)} \\ \hline
         $n$ & $p$& 1.00 &2.00 & 3.00 & 6.00 & 7.00& 8.00 & 9.00& 10.00 \\ \hline
10 & 0.10 & ( 0.87, 0.90, 0.86) & ( 0.88, 0.90, 0.86) & ( 0.88, 0.91, 0.87) & ( 0.90, 0.92, 0.89) & ( 0.91, 0.93, 0.89) & ( 0.91, 0.93, 0.90) & ( 0.92, 0.93, 0.91) & ( 0.92, 0.94, 0.91) \\
   & 0.25 & ( 0.93, 0.94, 0.91) & ( 0.93, 0.94, 0.92) & ( 0.94, 0.94, 0.93) & ( 0.96, 0.95, 0.95) & ( 0.96, 0.96, 0.96) & ( 0.97, 0.97, 0.96) & ( 0.97, 0.97, 0.96) & ( 0.97, 0.97, 0.97) \\
   & 0.50 & ( 1.00, 0.98, 1.00) & ( 1.00, 0.99, 1.00) & ( 1.00, 0.99, 1.00) & ( 1.00, 1.00, 1.00) & ( 1.00, 1.00, 1.00) & ( 1.00, 1.00, 1.00) & ( 1.01, 1.00, 1.00) & ( 1.02, 1.00, 1.00) \\
   & 0.75 & ( 1.02, 1.02, 1.02) & ( 1.03, 1.02, 1.03) & ( 1.03, 1.03, 1.03) & ( 1.04, 1.04, 1.05) & ( 1.05, 1.04, 1.05) & ( 1.05, 1.05, 1.05) & ( 1.05, 1.05, 1.05) & ( 1.05, 1.05, 1.06) \\
   & 0.90 & ( 1.06, 1.06, 1.06) & ( 1.06, 1.06, 1.06) & ( 1.06, 1.06, 1.07) & ( 1.07, 1.07, 1.08) & ( 1.07, 1.07, 1.08) & ( 1.07, 1.08, 1.08) & ( 1.08, 1.08, 1.09) & ( 1.08, 1.08, 1.09) \\ \hline
20 & 0.10 & ( 0.91, 0.93, 0.92) & ( 0.91, 0.93, 0.93) & ( 0.92, 0.94, 0.93) & ( 0.93, 0.95, 0.94) & ( 0.94, 0.95, 0.94) & ( 0.94, 0.95, 0.95) & ( 0.94, 0.95, 0.95) & ( 0.94, 0.96, 0.95) \\
   & 0.25 & ( 0.95, 0.96, 0.96) & ( 0.95, 0.96, 0.96) & ( 0.96, 0.96, 0.96) & ( 0.97, 0.97, 0.97) & ( 0.97, 0.98, 0.98) & ( 0.98, 0.98, 0.98) & ( 0.98, 0.98, 0.98) & ( 0.98, 0.98, 0.98) \\
   & 0.50 & ( 1.00, 0.99, 1.00) & ( 1.00, 0.99, 1.00) & ( 1.00, 1.00, 1.00) & ( 1.00, 1.00, 1.00) & ( 1.00, 1.00, 1.01) & ( 1.01, 1.00, 1.01) & ( 1.01, 1.00, 1.01) & ( 1.02, 1.00, 1.02) \\
   & 0.75 & ( 1.02, 1.02, 1.02) & ( 1.02, 1.02, 1.02) & ( 1.02, 1.02, 1.02) & ( 1.03, 1.03, 1.03) & ( 1.03, 1.03, 1.03) & ( 1.03, 1.03, 1.03) & ( 1.04, 1.04, 1.03) & ( 1.04, 1.04, 1.03) \\
   & 0.90 & ( 1.04, 1.04, 1.04) & ( 1.04, 1.04, 1.04) & ( 1.04, 1.04, 1.04) & ( 1.05, 1.05, 1.04) & ( 1.05, 1.05, 1.05) & ( 1.05, 1.06, 1.05) & ( 1.06, 1.06, 1.05) & ( 1.06, 1.06, 1.05) \\ \hline
30 & 0.10 & ( 0.93, 0.95, 0.93) & ( 0.93, 0.95, 0.94) & ( 0.93, 0.95, 0.94) & ( 0.94, 0.96, 0.95) & ( 0.95, 0.96, 0.95) & ( 0.95, 0.96, 0.96) & ( 0.95, 0.96, 0.96) & ( 0.95, 0.96, 0.96) \\
   & 0.25 & ( 0.96, 0.97, 0.96) & ( 0.96, 0.97, 0.97) & ( 0.96, 0.97, 0.97) & ( 0.98, 0.98, 0.98) & ( 0.98, 0.98, 0.98) & ( 0.98, 0.98, 0.98) & ( 0.98, 0.98, 0.99) & ( 0.99, 0.99, 0.99) \\
   & 0.50 & ( 1.00, 0.99, 1.00) & ( 1.00, 1.00, 1.00) & ( 1.00, 1.00, 1.00) & ( 1.00, 1.00, 1.00) & ( 1.00, 1.00, 1.00) & ( 1.01, 1.00, 1.01) & ( 1.01, 1.00, 1.01) & ( 1.02, 1.00, 1.01) \\
   & 0.75 & ( 1.02, 1.01, 1.01) & ( 1.02, 1.01, 1.02) & ( 1.02, 1.02, 1.02) & ( 1.03, 1.02, 1.02) & ( 1.03, 1.03, 1.02) & ( 1.03, 1.03, 1.03) & ( 1.03, 1.03, 1.03) & ( 1.03, 1.03, 1.03) \\
   & 0.90 & ( 1.03, 1.03, 1.03) & ( 1.04, 1.04, 1.03) & ( 1.04, 1.04, 1.03) & ( 1.04, 1.04, 1.04) & ( 1.04, 1.04, 1.04) & ( 1.04, 1.05, 1.04) & ( 1.05, 1.05, 1.04) & ( 1.05, 1.05, 1.04) \\ \hline
40 & 0.10 & ( 0.93, 0.95, 0.94) & ( 0.94, 0.95, 0.94) & ( 0.94, 0.96, 0.95) & ( 0.95, 0.96, 0.96) & ( 0.95, 0.96, 0.96) & ( 0.96, 0.97, 0.96) & ( 0.96, 0.97, 0.96) & ( 0.96, 0.97, 0.96) \\
   & 0.25 & ( 0.96, 0.97, 0.97) & ( 0.97, 0.97, 0.97) & ( 0.97, 0.97, 0.97) & ( 0.98, 0.98, 0.98) & ( 0.98, 0.98, 0.98) & ( 0.98, 0.99, 0.99) & ( 0.99, 0.99, 0.99) & ( 0.99, 0.99, 0.99) \\
   & 0.50 & ( 1.00, 0.99, 1.00) & ( 1.00, 1.00, 1.00) & ( 1.00, 1.00, 1.00) & ( 1.00, 1.00, 1.00) & ( 1.00, 1.00, 1.00) & ( 1.01, 1.00, 1.01) & ( 1.01, 1.00, 1.01) & ( 1.01, 1.00, 1.01) \\
   & 0.75 & ( 1.01, 1.01, 1.01) & ( 1.02, 1.01, 1.02) & ( 1.02, 1.01, 1.02) & ( 1.02, 1.02, 1.02) & ( 1.03, 1.02, 1.02) & ( 1.03, 1.02, 1.02) & ( 1.03, 1.03, 1.03) & ( 1.03, 1.03, 1.03) \\
   & 0.90 & ( 1.03, 1.03, 1.03) & ( 1.03, 1.03, 1.03) & ( 1.03, 1.03, 1.03) & ( 1.04, 1.04, 1.03) & ( 1.04, 1.04, 1.03) & ( 1.04, 1.04, 1.04) & ( 1.04, 1.04, 1.04) & ( 1.04, 1.04, 1.04) \\ \hline
50 & 0.10 & ( 0.94, 0.96, 0.94) & ( 0.94, 0.96, 0.95) & ( 0.95, 0.96, 0.95) & ( 0.96, 0.97, 0.96) & ( 0.96, 0.97, 0.96) & ( 0.96, 0.97, 0.96) & ( 0.96, 0.97, 0.97) & ( 0.96, 0.97, 0.97) \\
   & 0.25 & ( 0.97, 0.97, 0.97) & ( 0.97, 0.98, 0.97) & ( 0.97, 0.98, 0.97) & ( 0.98, 0.98, 0.98) & ( 0.98, 0.98, 0.98) & ( 0.99, 0.99, 0.99) & ( 0.99, 0.99, 0.99) & ( 0.99, 0.99, 0.99) \\
   & 0.50 & ( 1.00, 0.99, 1.00) & ( 1.00, 1.00, 1.00) & ( 1.00, 1.00, 1.00) & ( 1.00, 1.00, 1.00) & ( 1.00, 1.00, 1.00) & ( 1.01, 1.00, 1.01) & ( 1.01, 1.00, 1.01) & ( 1.01, 1.00, 1.01) \\
   & 0.75 & ( 1.01, 1.01, 1.01) & ( 1.01, 1.01, 1.01) & ( 1.02, 1.01, 1.02) & ( 1.02, 1.02, 1.02) & ( 1.02, 1.02, 1.02) & ( 1.02, 1.02, 1.02) & ( 1.03, 1.02, 1.02) & ( 1.03, 1.02, 1.02) \\
   & 0.90 & ( 1.03, 1.03, 1.03) & ( 1.03, 1.03, 1.03) & ( 1.03, 1.03, 1.03) & ( 1.03, 1.03, 1.03) & ( 1.03, 1.03, 1.03) & ( 1.04, 1.04, 1.03) & ( 1.04, 1.04, 1.04) & ( 1.04, 1.04, 1.04) \\ \hline
    \end{tabular}
   \end{adjustbox}{}}
\end{table}
\end{table}

\begin{table}[H]
\ContinuedFloat
		\caption{ (cont.): Goodness of $W(c, d)$ as the Approximate Distribution of $(\hat{\beta}/\beta)$}
	\begin{table}[H]\fontsize{9}{10}\selectfont
    \centering
    \rotatebox{90}{
   \begin{adjustbox}{max width=1.31\textwidth}
 \centering
    \begin{tabular}{|l|l|l|l|l|l|l|l|l|l|l|}
    \hline
         & &  \multicolumn{8}{|c|}{$\delta$} \\ 
         & &  \multicolumn{8}{|c|}{($\hat{Q}^*_p, ~\tilde{Q}^*_p$, ~$\hat {\hat {Q}}^*_p$)} \\ \hline
         $n$ & $p$& 1.00 &2.00 & 3.00 & 6.00 & 7.00& 8.00 & 9.00& 10.00 \\ \hline
60  & 0.10 & ( 0.95, 0.96, 0.95) & ( 0.95, 0.96, 0.95) & ( 0.95, 0.96, 0.95) & ( 0.96, 0.97, 0.96) & ( 0.96, 0.97, 0.96) & ( 0.96, 0.97, 0.97) & ( 0.97, 0.97, 0.97) & ( 0.97, 0.98, 0.97) \\
    & 0.25 & ( 0.97, 0.98, 0.97) & ( 0.97, 0.98, 0.97) & ( 0.97, 0.98, 0.98) & ( 0.98, 0.98, 0.98) & ( 0.99, 0.99, 0.99) & ( 0.99, 0.99, 0.99) & ( 0.99, 0.99, 0.99) & ( 0.99, 0.99, 0.99) \\
    & 0.50 & ( 1.00, 1.00, 1.00) & ( 1.00, 1.00, 1.00) & ( 1.00, 1.00, 1.00) & ( 1.00, 1.00, 1.00) & ( 1.00, 1.00, 1.00) & ( 1.01, 1.00, 1.01) & ( 1.01, 1.00, 1.01) & ( 1.01, 1.00, 1.01) \\
    & 0.75 & ( 1.01, 1.01, 1.01) & ( 1.01, 1.01, 1.01) & ( 1.01, 1.01, 1.02) & ( 1.02, 1.02, 1.02) & ( 1.02, 1.02, 1.02) & ( 1.02, 1.02, 1.02) & ( 1.02, 1.02, 1.02) & ( 1.02, 1.02, 1.02) \\
    & 0.90 & ( 1.02, 1.02, 1.02) & ( 1.03, 1.03, 1.03) & ( 1.03, 1.03, 1.03) & ( 1.03, 1.03, 1.03) & ( 1.03, 1.03, 1.03) & ( 1.03, 1.03, 1.03) & ( 1.03, 1.04, 1.03) & ( 1.04, 1.04, 1.03) \\ \hline
70  & 0.10 & ( 0.95, 0.96, 0.95) & ( 0.95, 0.97, 0.95) & ( 0.95, 0.97, 0.95) & ( 0.96, 0.97, 0.96) & ( 0.96, 0.97, 0.96) & ( 0.97, 0.97, 0.97) & ( 0.97, 0.98, 0.97) & ( 0.97, 0.98, 0.97) \\
    & 0.25 & ( 0.97, 0.98, 0.97) & ( 0.97, 0.98, 0.97) & ( 0.98, 0.98, 0.98) & ( 0.98, 0.99, 0.98) & ( 0.99, 0.99, 0.99) & ( 0.99, 0.99, 0.99) & ( 0.99, 0.99, 0.99) & ( 0.99, 0.99, 0.99) \\
    & 0.50 & ( 1.00, 1.00, 1.00) & ( 1.00, 1.00, 1.00) & ( 1.00, 1.00, 1.00) & ( 1.00, 1.00, 1.00) & ( 1.00, 1.00, 1.00) & ( 1.00, 1.00, 1.01) & ( 1.01, 1.00, 1.01) & ( 1.01, 1.00, 1.01) \\
    & 0.75 & ( 1.01, 1.01, 1.01) & ( 1.01, 1.01, 1.01) & ( 1.01, 1.01, 1.01) & ( 1.02, 1.02, 1.02) & ( 1.02, 1.02, 1.02) & ( 1.02, 1.02, 1.02) & ( 1.02, 1.02, 1.02) & ( 1.02, 1.02, 1.02) \\
    & 0.90 & ( 1.02, 1.02, 1.02) & ( 1.02, 1.02, 1.02) & ( 1.02, 1.02, 1.03) & ( 1.03, 1.03, 1.03) & ( 1.03, 1.03, 1.03) & ( 1.03, 1.03, 1.03) & ( 1.03, 1.03, 1.03) & ( 1.03, 1.03, 1.03) \\ \hline
80  & 0.10 & ( 0.95, 0.97, 0.95) & ( 0.96, 0.97, 0.95) & ( 0.96, 0.97, 0.96) & ( 0.96, 0.97, 0.96) & ( 0.97, 0.98, 0.97) & ( 0.97, 0.98, 0.97) & ( 0.97, 0.98, 0.97) & ( 0.97, 0.98, 0.97) \\
    & 0.25 & ( 0.97, 0.98, 0.97) & ( 0.98, 0.98, 0.98) & ( 0.98, 0.98, 0.98) & ( 0.99, 0.99, 0.98) & ( 0.99, 0.99, 0.99) & ( 0.99, 0.99, 0.99) & ( 0.99, 0.99, 0.99) & ( 0.99, 0.99, 0.99) \\
    & 0.50 & ( 1.00, 1.00, 1.00) & ( 1.00, 1.00, 1.00) & ( 1.00, 1.00, 1.00) & ( 1.00, 1.00, 1.00) & ( 1.00, 1.00, 1.00) & ( 1.00, 1.00, 1.01) & ( 1.01, 1.00, 1.01) & ( 1.01, 1.00, 1.01) \\
    & 0.75 & ( 1.01, 1.01, 1.01) & ( 1.01, 1.01, 1.01) & ( 1.01, 1.01, 1.01) & ( 1.02, 1.02, 1.02) & ( 1.02, 1.02, 1.02) & ( 1.02, 1.02, 1.02) & ( 1.02, 1.02, 1.02) & ( 1.02, 1.02, 1.02) \\
    & 0.90 & ( 1.02, 1.02, 1.02) & ( 1.02, 1.02, 1.02) & ( 1.02, 1.02, 1.02) & ( 1.03, 1.03, 1.03) & ( 1.03, 1.03, 1.03) & ( 1.03, 1.03, 1.03) & ( 1.03, 1.03, 1.03) & ( 1.03, 1.03, 1.03) \\ \hline
90  & 0.10 & ( 0.96, 0.97, 0.95) & ( 0.96, 0.97, 0.96) & ( 0.96, 0.97, 0.96) & ( 0.97, 0.97, 0.97) & ( 0.97, 0.98, 0.97) & ( 0.97, 0.98, 0.97) & ( 0.97, 0.98, 0.97) & ( 0.97, 0.98, 0.97) \\
    & 0.25 & ( 0.98, 0.98, 0.97) & ( 0.98, 0.98, 0.98) & ( 0.98, 0.98, 0.98) & ( 0.99, 0.99, 0.99) & ( 0.99, 0.99, 0.99) & ( 0.99, 0.99, 0.99) & ( 0.99, 0.99, 0.99) & ( 0.99, 0.99, 0.99) \\
    & 0.50 & ( 1.00, 1.00, 1.00) & ( 1.00, 1.00, 1.00) & ( 1.00, 1.00, 1.00) & ( 1.00, 1.00, 1.00) & ( 1.00, 1.00, 1.00) & ( 1.00, 1.00, 1.01) & ( 1.01, 1.00, 1.01) & ( 1.01, 1.00, 1.01) \\
    & 0.75 & ( 1.01, 1.01, 1.01) & ( 1.01, 1.01, 1.01) & ( 1.01, 1.01, 1.01) & ( 1.02, 1.01, 1.02) & ( 1.02, 1.02, 1.02) & ( 1.02, 1.02, 1.02) & ( 1.02, 1.02, 1.02) & ( 1.02, 1.02, 1.02) \\
    & 0.90 & ( 1.02, 1.02, 1.02) & ( 1.02, 1.02, 1.02) & ( 1.02, 1.02, 1.02) & ( 1.02, 1.03, 1.03) & ( 1.03, 1.03, 1.03) & ( 1.03, 1.03, 1.03) & ( 1.03, 1.03, 1.03) & ( 1.03, 1.03, 1.03) \\ \hline
100 & 0.10 & ( 0.96, 0.97, 0.95) & ( 0.96, 0.97, 0.96) & ( 0.96, 0.97, 0.96) & ( 0.97, 0.98, 0.97) & ( 0.97, 0.98, 0.97) & ( 0.97, 0.98, 0.97) & ( 0.97, 0.98, 0.97) & ( 0.97, 0.98, 0.97) \\
    & 0.25 & ( 0.98, 0.98, 0.98) & ( 0.98, 0.98, 0.98) & ( 0.98, 0.98, 0.98) & ( 0.99, 0.99, 0.99) & ( 0.99, 0.99, 0.99) & ( 0.99, 0.99, 0.99) & ( 0.99, 0.99, 0.99) & ( 0.99, 0.99, 0.99) \\
    & 0.50 & ( 1.00, 1.00, 1.00) & ( 1.00, 1.00, 1.00) & ( 1.00, 1.00, 1.00) & ( 1.00, 1.00, 1.00) & ( 1.00, 1.00, 1.00) & ( 1.00, 1.00, 1.01) & ( 1.01, 1.00, 1.01) & ( 1.01, 1.00, 1.01) \\
    & 0.75 & ( 1.01, 1.01, 1.01) & ( 1.01, 1.01, 1.01) & ( 1.01, 1.01, 1.01) & ( 1.02, 1.01, 1.02) & ( 1.02, 1.01, 1.02) & ( 1.02, 1.02, 1.02) & ( 1.02, 1.02, 1.02) & ( 1.02, 1.02, 1.02) \\
    & 0.90 & ( 1.02, 1.02, 1.02) & ( 1.02, 1.02, 1.02) & ( 1.02, 1.02, 1.02) & ( 1.02, 1.02, 1.03) & ( 1.02, 1.03, 1.03) & ( 1.03, 1.03, 1.03) & ( 1.03, 1.03, 1.03) & ( 1.03, 1.03, 1.03) \\ \hline
   \end{tabular}
   \end{adjustbox}{}}
\end{table}
\end{table}

\begin{figure}[H]
    \includegraphics[width=15cm]{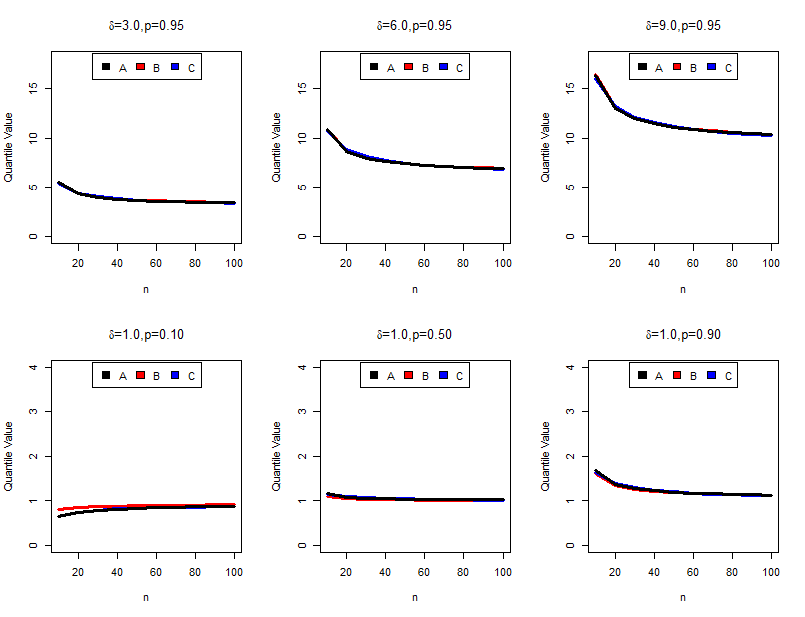}
    \centering
	\caption{Plots of $\hat{Q}_p$(= A)$, ~\tilde{Q}_p$(= B) and $\hat {\hat {Q}}_p(= C)$  as functions of $n$}

    \includegraphics[width=15cm]{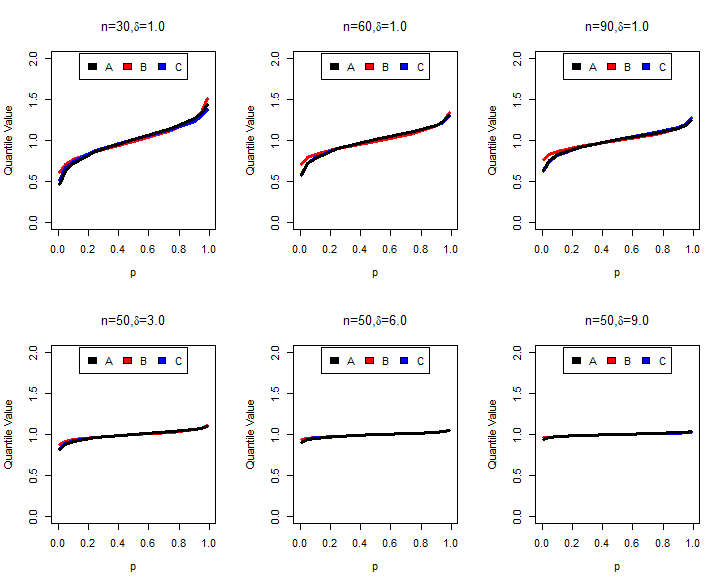}
    \centering
	\caption{Plots of $\hat{Q}^*_p$(= A)$, ~\tilde{Q}^*_p$(= B) and $\hat {\hat {Q}}^*_p(= C)$ as functions of $p$}
\end{figure}

\begin{figure}[H]
    \includegraphics[width=15cm]{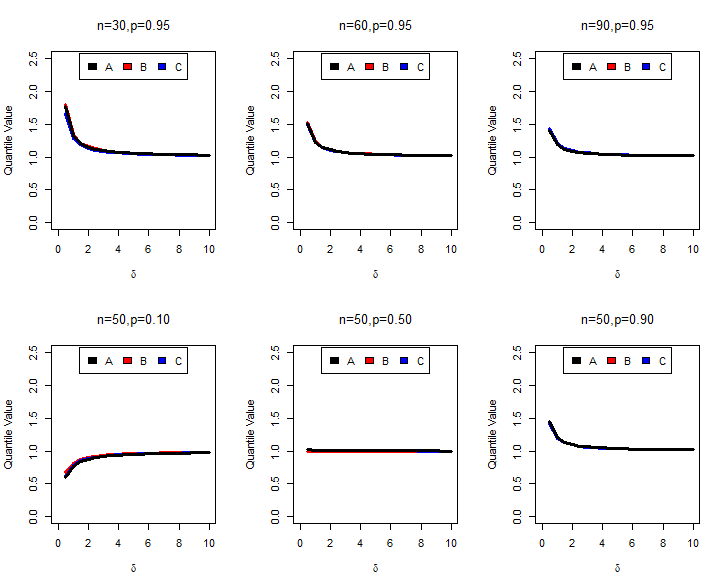}
    \centering
	\caption{Plots of $\hat{Q}^*_p$(= A)$, ~\tilde{Q}^*_p$(= B) and $\hat {\hat {Q}}^*_p(= C)$  as functions of $\delta$}

    \includegraphics[width=15cm]{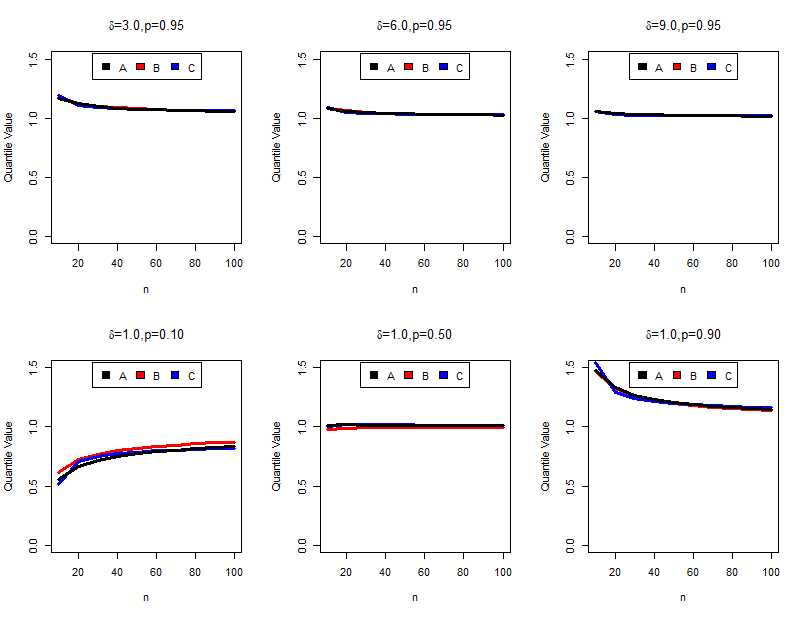}
	\caption{Plots of $\hat{Q}^*_p$(= A)$, ~\tilde{Q}^*_p$(= B) and $\hat {\hat {Q}}^*_p(= C)$  as functions of $n$}
\end{figure}

It should be kept in mind that, in asymptotic theory when we use the first order principle, the mean squared error (MSE) is equivalent to the variance, since $\text{MSE}=\text{Var}+(\text{bias})^2$, and $(\text{bias})^2$ contributes a term of order $O(n^{-2})$ which is considered negligible. That is why we are using the term FOV, and not the \lq first order MSE\rq.

On the other hand, in Section 3 we have hypothesized that $\hat{\delta}~\dot{\sim}~ W(a,b) \text{ and } (\hat\beta/\beta) ~\dot{\sim}~ W(c,d)$. Based on these approximate sampling distributions 
\begin{equation}
    \text{Bias}(\hat{\delta})\approx b \Gamma(1+1/a)-\delta=(1/n)[n\{b\Gamma(1+1/a)-\delta\}]\approx(\hat{B_1}/n) \text{~~(say)}, 
\end{equation}
where
\begin{equation}
  \hat{B_1}=n\{\hat{b}\Gamma(1+1/\hat{a})-\delta\},  
\end{equation}
with $\hat{a}=\hat{a}_0+\hat{a}_1\textit{ln}{(n+\hat{a}_2)}$ and $\hat{b}=\hat{b}_0/\delta\{1+\hat{b}_1\textit{ln}{(\textit{ln}(n))}\}$. Also, 
\begin{equation}
\begin{split}
  \text{Var}(\hat{\delta})&\approx b^2[\Gamma(1+2/a)-\{\Gamma(1+1/a)\}^2]\\
  &=(1/n)[nb^2\{\Gamma(1+2/a)-(\Gamma(1+1/a))^2\}]\approx(\hat{V_1}/n)  \text{~(say)},
\end{split}
\end{equation}
where
\begin{equation}
    \hat{V_1}=n\hat{b}^2\{\Gamma(1+2/\hat{a})-(\Gamma(1+1/\hat{a}))^2\}.  
\end{equation}

Exactly in a similar manner, we can write
\begin{equation}
\begin{split}
    \text{Bias}(\hat{\beta}/\beta)&\approx d\Gamma(1+1/c)-1\\
    &=(1/n)[n\{d\Gamma(1+1/c)-1\}]\\
    &\approx(\hat{B_2}/n) \text{~~(say)}, 
\end{split}
\end{equation}
 where
 \begin{equation}
     \hat{B_2}=n\{\hat{d}\Gamma(1+1/\hat{c})-1\}, 
 \end{equation}
and 
\begin{equation}
\begin{split}
    \text{Var}(\hat{\beta}/\beta) &\approx d^2[\Gamma(1+2/c)-\{\Gamma(1+1/c)\}^2]\\
    &=(1/n)[nd^2\{\Gamma(1+2/c)-(\Gamma(1+1/c))^2\}]
    \approx(\hat{V_2}/n) \text{~(say)},
    \end{split}
\end{equation}
where
\begin{equation}
    \hat{V_2}=n\hat{d}^2\{\Gamma(1+2/\hat{c})-(\Gamma(1+1/\hat{c}))^2\}, 
\end{equation}
with $\hat{c}=\hat{c}_0\delta\textit{ln}(n+\hat{c}_1)$ and $\hat{d}=\hat{d}_0+\hat{d}_1/n+\hat{d}_2/\delta+\hat{d}_3/n\delta.$

All the estimated coefficients of $\hat{a}, \hat{b}, \hat{c}$ and $\hat{d} ~($i.e, $(\hat{a}_0,\hat{a}_1,\hat{a}_2),(\hat{b}_0,\hat{b}_1),(\hat{c}_0,\hat{c}_1)$ and $(\hat{d}_0,\hat{d}_1,\hat{d}_2,\hat{d}_3))$ have been provided in Table $\ref{tab5}$.

Our objective here is to compare (i) $B_1$ with $\hat{B_1}$; (ii) $V_1$ with $\hat{V_1}$; (iii) $B_2$ with $\hat{B_2}$; and (iv) $V_2$ with $\hat{V_2}$. Note that all these first order terms are functions of $(n,\delta)$, however, $B_1, V_1, B_2$ and $V_2$ are functions of $\delta$ only.

The following Tables \ref{part3:1} $-$ \ref{part3:4} show the aforementioned comparisons where the computed values of $|\hat{B_1}-B_1|/n, ~|\hat{V_1}-V_1|/n, ~|\hat{B_2}-B_2|/n$ and $|\hat{V_2}-V_2|/n$ has been presented for various combinations of $(n,\delta)$.

\noindent {\bf{Remark 4.3}}. Tables $\ref{part3:1} - \ref{part3:4}$ actually show how good the FOB and FOV are compared to $(\hat{B}_1/n)$ and $(\hat{B}_2/n)$ as well as $(\hat{V}_1/n)$ and $(\hat{V}_2/n)$, respectively. Note that $(\hat{B}_i/n)$ and $(\hat{V}_i/n), ~i=1,2,$ are more accurate as they are based on the hypothesized Weibull distributions as shown in Section 3. As expected, FOB and FOV values catch up with $(\hat{B}_i/n)$ and $(\hat{V}_i/n)$ for $n\geq 50$ for $\hat{\delta}$, and for $n\geq 10$ for $(\hat{\beta}/\beta)$, irrespective of the value of $\delta$. Note that FOB of $(\hat{\beta}/\beta)$ is very close to the actual Bias$(\hat{\beta}/\beta)$, and FOV of $(\hat{\beta}/\beta)$ is very close to the actual Var$(\hat{\beta}/\beta)$ as seen from Tables 4.5 and 4.6, respectively, especially of order 0.01 for sample sizes greater than 20. On the other hand, it may seem that, in an absolute sense, FOB of $\hat{\delta}$ as well as FOV of $\hat{\delta}$ are not converging fast enough to Bias($\hat{\delta}$) and Var($\hat{\delta}$) as $n$ increases. But in a relative sense, i.e., when the actual Bias, as well as FOB, are divided by $\delta$, and the actual Var, as well as  FOV, are divided by $\delta^2$, then in terms of \lq relative bias\rq and \lq relative variance\rq \, these convergences are comparable to those of $(\hat{\beta}/\beta)$. 

\begin{table}[H]\fontsize{10}{10}\selectfont
		\caption{Computation of the absolute difference between $\hat{B_1}/n$ and $B_1/n$ for $\hat{\delta}$}
		\label{part3:1}
\begin{adjustbox}{max width=\textwidth}
	\centering
    \begin{tabular}{c|l|l|l|l|l|l|l|l|l|l|l|l}
    \hline
		$|\hat{B_1}-B_1|/n$& \multicolumn{12}{|c}{$\delta$}    \\ \hline
        $n$ & 0.5 & 1.0 & 1.5 & 2.0 & 3.0 & 4.0 & 5.0 & 6.0 & 7.0 & 8.0 & 9.0 & 10.0 \\ \hline
        10 & 0.012 & 0.024 & 0.037 & 0.048 & 0.074 & 0.093 & 0.113 & 0.141 & 0.177 & 0.191 & 0.215 & 0.243 \\ 
        20 & 0.001 & 0.000 & 0.003 & 0.003 & 0.003 & 0.004 & 0.015 & 0.005 & 0.006 & 0.016 & 0.019 & 0.008 \\ 
        30 & 0.003 & 0.005 & 0.006 & 0.010 & 0.016 & 0.018 & 0.025 & 0.026 & 0.034 & 0.041 & 0.047 & 0.042 \\ 
        40 & 0.002 & 0.006 & 0.008 & 0.012 & 0.016 & 0.023 & 0.027 & 0.029 & 0.033 & 0.050 & 0.055 & 0.052 \\ 
        50 & 0.003 & 0.005 & 0.009 & 0.012 & 0.018 & 0.025 & 0.029 & 0.033 & 0.040 & 0.043 & 0.054 & 0.053 \\ 
        60 & 0.003 & 0.005 & 0.008 & 0.011 & 0.017 & 0.021 & 0.028 & 0.033 & 0.042 & 0.044 & 0.046 & 0.052 \\ 
        70 & 0.003 & 0.006 & 0.009 & 0.009 & 0.016 & 0.018 & 0.027 & 0.033 & 0.036 & 0.042 & 0.051 & 0.053 \\ 
        80 & 0.003 & 0.005 & 0.008 & 0.010 & 0.015 & 0.022 & 0.030 & 0.029 & 0.035 & 0.046 & 0.044 & 0.053 \\ 
        90 & 0.003 & 0.005 & 0.007 & 0.010 & 0.016 & 0.022 & 0.027 & 0.031 & 0.034 & 0.034 & 0.045 & 0.048 \\ 
        100 & 0.002 & 0.005 & 0.008 & 0.010 & 0.014 & 0.018 & 0.025 & 0.029 & 0.031 & 0.034 & 0.043 & 0.051 \\ \hline
    \end{tabular}
\end{adjustbox}
\end{table}

\begin{table}[H]\fontsize{10}{10}\selectfont
   \caption{Computation of the absolute difference between $\hat{V_1}/n$ and $V_1/n$ for $\hat{\delta}$}
    \label{part3:2}
\begin{adjustbox}{max width=\textwidth}
	\centering
    \begin{tabular}{c|l|l|l|l|l|l|l|l|l|l|l|l}
    \hline
		$|\hat{V_1}-V_1|/n$& \multicolumn{12}{|c}{$\delta$}    \\ \hline
        $n$ & 0.5 & 1.0 & 1.5 & 2.0 & 3.0 & 4.0 & 5.0 & 6.0 & 7.0 & 8.0 & 9.0 & 10.0 \\ \hline
        10 & 0.023 & 0.095 & 0.209 & 0.375 & 0.843 & 1.481 & 2.320 & 3.374 & 4.648 & 6.119 & 7.615 & 9.495 \\ 
        20 & 0.007 & 0.028 & 0.065 & 0.110 & 0.259 & 0.461 & 0.707 & 1.013 & 1.382 & 1.827 & 2.298 & 2.724 \\ 
        30 & 0.004 & 0.016 & 0.035 & 0.065 & 0.143 & 0.258 & 0.397 & 0.583 & 0.784 & 1.033 & 1.299 & 1.630 \\ 
        40 & 0.003 & 0.011 & 0.026 & 0.042 & 0.100 & 0.174 & 0.279 & 0.398 & 0.534 & 0.683 & 0.928 & 1.124 \\ 
        50 & 0.002 & 0.009 & 0.019 & 0.034 & 0.076 & 0.139 & 0.218 & 0.311 & 0.416 & 0.556 & 0.701 & 0.852 \\ 
        60 & 0.002 & 0.007 & 0.015 & 0.027 & 0.062 & 0.113 & 0.169 & 0.248 & 0.344 & 0.442 & 0.540 & 0.691 \\ 
        70 & 0.001 & 0.006 & 0.013 & 0.022 & 0.053 & 0.090 & 0.145 & 0.202 & 0.278 & 0.359 & 0.461 & 0.564 \\ 
        80 & 0.001 & 0.005 & 0.011 & 0.020 & 0.042 & 0.078 & 0.126 & 0.177 & 0.243 & 0.320 & 0.392 & 0.492 \\ 
        90 & 0.001 & 0.004 & 0.010 & 0.018 & 0.040 & 0.071 & 0.108 & 0.157 & 0.211 & 0.279 & 0.351 & 0.433 \\ 
        100 & 0.001 & 0.004 & 0.009 & 0.015 & 0.035 & 0.063 & 0.095 & 0.136 & 0.194 & 0.246 & 0.312 & 0.388 \\ \hline
    \end{tabular}
\end{adjustbox}
\end{table}

\begin{table}[H]\fontsize{10}{10}\selectfont
   \caption{Computation of the absolute difference between $\hat{B_2}/n$ and $B_2/n$ for $(\hat{\beta}/\beta)$}
    \label{part3:3}
\begin{adjustbox}{max width=\textwidth}
	\centering
    \begin{tabular}{c|l|l|l|l|l|l|l|l|l|l|l|l}
    \hline
		$|\hat{B_2}-B_2|/n$& \multicolumn{12}{|c}{$\delta$}    \\ \hline
        n & 0.5 & 1.0 & 1.5 & 2.0 & 3.0 & 4.0 & 5.0 & 6.0 & 7.0 & 8.0 & 9.0 & 10.0 \\ \hline
        10 & 0.004 & 0.005 & 0.004 & 0.004 & 0.003 & 0.003 & 0.003 & 0.002 & 0.002 & 0.002 & 0.002 & 0.002 \\
        20 & 0.005 & 0.004 & 0.005 & 0.004 & 0.003 & 0.003 & 0.002 & 0.002 & 0.002 & 0.001 & 0.001 & 0.001 \\ 
        30 & 0.002 & 0.004 & 0.004 & 0.003 & 0.004 & 0.002 & 0.002 & 0.002 & 0.001 & 0.001 & 0.001 & 0.001 \\ 
        40 & 0.003 & 0.004 & 0.004 & 0.004 & 0.003 & 0.002 & 0.002 & 0.002 & 0.001 & 0.001 & 0.001 & 0.001 \\ 
        50 & 0.003 & 0.006 & 0.004 & 0.004 & 0.003 & 0.002 & 0.002 & 0.001 & 0.001 & 0.001 & 0.001 & 0.001 \\ 
        60 & 0.006 & 0.005 & 0.004 & 0.003 & 0.002 & 0.002 & 0.001 & 0.001 & 0.001 & 0.001 & 0.001 & 0.001 \\ 
        70 & 0.002 & 0.005 & 0.003 & 0.003 & 0.002 & 0.002 & 0.001 & 0.001 & 0.001 & 0.001 & 0.001 & 0.001 \\ 
        80 & 0.005 & 0.004 & 0.004 & 0.003 & 0.002 & 0.002 & 0.001 & 0.001 & 0.001 & 0.001 & 0.001 & 0.001 \\ 
        90 & 0.005 & 0.004 & 0.003 & 0.003 & 0.002 & 0.002 & 0.001 & 0.001 & 0.001 & 0.001 & 0.001 & 0.001 \\ 
        100 & 0.004 & 0.004 & 0.003 & 0.003 & 0.002 & 0.002 & 0.001 & 0.001 & 0.001 & 0.001 & 0.001 & 0.001 \\ \hline
    \end{tabular}
\end{adjustbox}
\end{table}

\begin{table}[H]\fontsize{10}{10}\selectfont
   \caption{Computation of the absolute difference between $\hat{V_2}/n$ and $V_2/n$ for $(\hat{\beta}/\beta)$}
    \label{part3:4}
\begin{adjustbox}{max width=\textwidth}
	\centering
    \begin{tabular}{c|l|l|l|l|l|l|l|l|l|l|l|l}
    \hline
		$|\hat{V_2}-V_2|/n$& \multicolumn{12}{|c}{$\delta$}    \\ \hline
        $n$ & 0.5 & 1.0 & 1.5 & 2.0 & 3.0 & 4.0 & 5.0 & 6.0 & 7.0 & 8.0 & 9.0 & 10.0 \\ \hline
        10 & 0.116 & 0.011 & 0.021 & 0.026 & 0.029 & 0.031 & 0.031 & 0.032 & 0.032 & 0.033 & 0.033 & 0.033 \\ 
        20 & 0.045 & 0.001 & 0.009 & 0.011 & 0.014 & 0.015 & 0.016 & 0.016 & 0.016 & 0.016 & 0.016 & 0.016 \\ 
        30 & 0.029 & 0.001 & 0.005 & 0.007 & 0.009 & 0.010 & 0.010 & 0.011 & 0.011 & 0.011 & 0.011 & 0.011 \\ 
        40 & 0.026 & 0.002 & 0.003 & 0.005 & 0.007 & 0.007 & 0.008 & 0.008 & 0.008 & 0.008 & 0.008 & 0.008 \\ 
        50 & 0.021 & 0.002 & 0.002 & 0.004 & 0.005 & 0.006 & 0.006 & 0.006 & 0.006 & 0.006 & 0.006 & 0.007 \\ 
        60 & 0.019 & 0.002 & 0.002 & 0.003 & 0.004 & 0.005 & 0.005 & 0.005 & 0.005 & 0.005 & 0.005 & 0.005 \\ 
        70 & 0.016 & 0.002 & 0.001 & 0.003 & 0.004 & 0.004 & 0.004 & 0.004 & 0.005 & 0.005 & 0.005 & 0.005 \\ 
        80 & 0.015 & 0.002 & 0.001 & 0.002 & 0.003 & 0.004 & 0.004 & 0.004 & 0.004 & 0.004 & 0.004 & 0.004 \\ 
        90 & 0.014 & 0.002 & 0.001 & 0.002 & 0.003 & 0.003 & 0.003 & 0.003 & 0.004 & 0.004 & 0.004 & 0.004 \\ 
        100 & 0.013 & 0.002 & 0.001 & 0.002 & 0.003 & 0.003 & 0.003 & 0.003 & 0.003 & 0.003 & 0.003 & 0.003 \\ \hline
    \end{tabular}
\end{adjustbox}
\end{table}
\section{Concluding Remarks}
\begin{itemize}
 \item[(a)] The main contribution of this work has been to show that the MLEs of the Weibull parameters 
approximately follow suitable Weibull distributions of their own, apart from proving analytically 
that these MLEs exist uniquely. Various evaluations have been undertaken to assess the goodness 
of these Weibull approximations to the actual simulated sampling distributions.

\item[(b)] We have also used the first-order bias and variance concept to measure the closeness of the
asymptotic distributions to their hypothesized sampling counterparts. As a consequence, this
study can help not only in estimating the Weibull percentiles but also in further inferences on $\delta$
and $\beta$, such as interval estimation of the model parameters as well as hypothesis testing which
might be more relevant from an applied point of view. The results of our subsequent inferential
aspects will be reported in the near future. 
\item[(c)] Another potential direction to extend our computational results is to consider censored observations
(under both Type-I and Type-II) and study the behavior of the model parameter MLEs in terms of
their sampling distributions. Definitely, it will be more complicated as the censoring regimen will play
a decisive role.
\item[(d)] Last but not least, this work is going to be useful mostly for small to moderately large sample 
sizes because the asymptotic normality holds comfortably for the MLEs for large sample sizes, and that 
is the reason (and as seen from Tables 4.3 – 4.6) why we have restricted our study to sample size up to 100.
\end{itemize}


\section*{Acknowledgment:} This work is the result of a short graduate-level summer (2023) research course taught by the third author at Ton Duc Thang University (TDTU), Ho Chi Minh City, Vietnam, from June 30 - July 22, 2023. The authors would like to thank the TDTU administration for their logistical support and hospitality.

\newpage
\section*{References}
\begin{description}

 \item  Almeida, J. B. (1999). Application of Weibull Statistics to the Failure of Coatings, \textit{Journal of Materials Processing Technology}, 93, 257–263.

\item Aljeddani, S. M. A. and Mohammed, M. A. (2023). A Novel Approach to Weibull Distribution for the Assessment of 
Wind Energy Speed, \textit{Alexandria Engineering Journal}, 78,  56 - 64. 

\item Austin, L. G., Klimpel, R. R., \& Luckie, P. T. (1984). \textit {Process engineering of size reduction: ball milling}. Society of Mining Engineers of the AIME.

\item Carroll, K. J. (2003). On the Use and utility of the Weibull Model in the Analysis of Survival Data, \textit{Controlled Clinical Trials}, 24, 682–701.

\item Chandrasekhar, S. (1943). Stochastic problems in physics and astronomy. \textit{Reviews of modern physics}, 15(1), 1.

\item Chatfield, C., \& Goodhardt, G. J. (1973). A consumer purchasing model with Erlang inter-purchase times. \textit{Journal of the American Statistical Association}, 68(344), 828-835.
\item Chen, M., Zhang, Z., \& Cui, C. (2017). On the bias of the maximum likelihood estimators of parameters of the Weibull distribution. \textit{Mathematical and Computational Applications}, 22(1), 19.

\item Collett, D. (2023). \textit{Modelling survival data in medical research}. Chapman and Hall/CRC.

\item Eliazar, I. (2017). Lindy’s law. \textit{Physica A: Statistical Mechanics and its Applications}, 486, 797-805. 

\item Cox, D. R., \& Snell, E. J. (1968). A general definition of residuals. \textit{Journal of the Royal Statistical Society: Series B (Methodological)}, 30(2), 248-265.

\item Fok, S. L., Mitchell, B. C., Smart, J. and Marsden, B. J. (2001). A Numerical Study on the Application of the Weibull Theory to Brittle Materials, \textit{Engineering Fracture Mechanics}, 68, 1171–1179.

\item Fleming, R. A. (2001). The Weibull Model and an Ecological Application: Describing the Dynamics of Foliage Biomass on Scots Pine, \textit{Ecological Modelling}, 138 (1-3), 309–319.

\item Heo, J. H., Boes, D. C. and Salas, J. D. (2001). Regional Flood Frequency Analysis Based on a Weibull Model: Part 1. Estimation, Asymptotic Variances, \textit{Journal of Hydrology}, 242, 157–170.

\item Johnson, N. L., Kotz, S. and Balakrishnan, N. (1994). \textit{Continuous Univariate Distributions}, Vol. 1 (Wiley Series in Probability and Statistics) 2nd Edition.

\item Jiang, R., \& Murthy, D. N. P. (2011). A study of Weibull shape parameter: Properties and significance. \textit{Reliability Engineering \& System Safety}, 96(12), 1619-1626.

\item Karlin, S. (1958). Admissibility for estimation with quadratic loss. \textit{The Annals of Mathematical Statistics}, 29(2), 406-436.

\item  Keshavan, K., Sargent, G. and Conrad, H. (1980). Statistical Analysis of the Hertzian Fracture of Pyrex Glass Using the Weibull Distribution  Function, \textit{Journal of Materials Science}, 15, 839–844.

\item Li, Q. S., Fang, J. Q., Liu, D. K. and Tang, J. (2003). Failure Probability Prediction of Concrete Components, \textit{Cement and Concrete Research}, 33, 1631–1636.

\item Mafart, P., Couvert, O., Gaillard, S. and Leguerinel, I. (2002). On Calculating Sterility in Thermal Preservation Methods: Application of the Weibull Frequency Distribution Model, \textit{International Journal of Food Microbiology}, 72, 107–113.

\item Makalic, E., \& Schmidt, D. F. (2023). Maximum likelihood estimation of the Weibull distribution with reduced bias. \textit{Statistics and Computing}, 33(3), 69.

\item Matsushita, S., Hagiwara, K., Shiota, T., Shimada, H., Kuramoto, K. and Toyokura, Y. (1992). Lifetime Data Analysis of Disease and Aging by the Weibull Probability Distribution, \textit{Journal of Clinical Epidemiology}, Vol. 45, No. 10, 1165-1175.

\item McCool, J. I. (2012). \textit {Using the Weibull distribution: reliability, modeling, and inference} (Vol. 950). John Wiley \& Sons, Inc., Hoboken, New Jersey.

\item Murthy, D. N. P., Bulmer, M. and Eccleston, J. E. (2004). Weibull modelling, \textit{Reliability Engineering and System Safety}, 86, 257–267.

\item Sharif, M. N., \& Islam, M. N. (1980). The Weibull distribution as a general model for forecasting technological change. \textit{Technological Forecasting and Social Change}, 18(3), 247-256.

\item Sheikh, A. K., Boah, J. K. and Hansen, D. A. (1990). Statistical Modelling of Pitting Corrosion and Pipeline Reliability, \textit{Corrosion}, 46, 190–196.

\item Tadikamalla, P. R. (1980). Age Replacement Policies for Weibull Failure Rates, \textit{IEEE Transactions on Reliability}, 29, 88–90.

\item Takeuchi, K., \& Akahira, M. (1979). Asymptotic optimality of the generalized Bayes estimator in multiparameter cases. \textit{Annals of the Institute of Statistical Mathematics}, 31, 403-415.

\item Tanaka, H., Pal, N., \& Lim, W. K. (2018). On improved estimation under Weibull model. \textit{Journal of Statistical Theory and Practice}, 12, 48-65.

\item Xie, M., Lai, C. D. and Murthy, D. N. P. (2003). Weibull-related distributions for modelling of bathtub shaped failure rate functions. 
In: \textit{Mathematical and Statistical Methods in Reliability}, pp.283–297, ed. by B.H. Lindqvist, K.A. Doksum, World Scientific, Singapore.

\end{description}

\end{document}